\documentclass[twocolumn]{aastex63}
\usepackage{amsmath}
\usepackage{comment}

\usepackage{enumitem}

\usepackage{physics}
\usepackage{CJK}

\newcommand{\be}{\begin{equation}}
\newcommand{\ee}{\end{equation}}

\shorttitle{Infrared Echoes from Dust-Shrouded Luminous Transients}
\shortauthors{Semih Tuna, B.~D.~Metzger}

\graphicspath{{./}}

\begin{document}
\begin{CJK*}{UTF8}{gbsn}

\title{Time-Dependent Radiation Transport Simulations of Infrared Echoes from Dust-Shrouded Luminous Transients}

\author[0000-0002-2002-6860]{Semih Tuna}
\affil{Department of Physics and Columbia Astrophysics Laboratory, Columbia University, New York, NY 10027, USA}

\author[0000-0002-4670-7509]{Brian D.~Metzger}
\affil{Department of Physics and Columbia Astrophysics Laboratory, Columbia University, New York, NY 10027, USA}
\affil{Center for Computational Astrophysics, Flatiron Institute, 162 5th Ave, New York, NY 10010, USA}

\author[0000-0002-2624-3399]{Yan-fei Jiang (姜燕飞)}
\affil{Center for Computational Astrophysics, Flatiron Institute, 162 5th Ave, New York, NY 10010, USA}

\author[0000-0001-7448-4253]{Christopher White}
\affil{Center for Computational Astrophysics, Flatiron Institute, 162 5th Ave, New York, NY 10010, USA}

\correspondingauthor{Semih Tuna}
\email{semih.tuna@columbia.edu}

\begin{abstract}

A wide range of stellar explosions, including supernovae (SNe), tidal disruption events (TDE), and fast blue optical transients (FBOT), can occur in dusty environments initially opaque to the transient's optical/UV light, becoming visible only once the dust is destroyed by the transient's rising luminosity. We present axisymmetric time-dependent radiation transport simulations of dust-shrouded transients with \texttt{Athena++} and tabulated gray opacities, which predict the light-curves of the dust-reprocessed infrared (IR) radiation. The luminosity and timescale of the IR light-curve depends on whether the transient rises rapidly or slowly compared to the light crossing-time of the photosphere, $t_{\rm lc}$.  For slow-rising transients ($t_{\rm rise} \gg t_{\rm lc}$) such as SNe, the reprocessed IR radiation diffuses outwards through the dust shell faster than the sublimation front expands; the IR light-curve therefore begins rising prior to the escape of UV/optical light, but peaks on a timescale $\sim t_{\rm rise}$ shorter than the transient duration. By contrast, for fast-rising transients ($t_{\rm rise} \ll t_{\rm lc}$) such as FBOTs and some TDEs, the finite light-travel time results in the reprocessed radiation arriving as an ``echo'' lasting much longer than the transient itself (despite the dust photosphere having already being destroyed by peak light).  We explore the effects of the system geometry by considering a torus-shaped distribution of dust. The IR light-curves seen by observers in the equatorial plane of the torus resemble those for a spherical dust shell, while polar observers see faster-rising, brighter and shorter-lived emission. We successfully model the IR excess seen in AT2018cow as a dust echo, supporting the presence of an opaque dusty medium surrounding FBOTs prior to explosion.
\end{abstract}


\section{Introduction}
\label{sec:introduction}

Optical and UV time-domain surveys are mapping out an ever growing ``zoo'' of explosive astrophysical transients. These include new classes of core-collapse supernovae (SNe), tidal disruptions events (TDE) of stars by massive black holes \citep{Rees88,Evans&Kochanek89}, luminous red novae (LRN) from stellar mergers \citep{Soker&Tylenda06,Tylenda+11,Ivanova+13}, as well as events of still-debated origin such as the ``intermediate luminosity red transients'' (ILRTs; \citealt{Bond+09,Thompson+09,Berger+09,Segev+19,Jencson+19}) and ``fast blue optical transients'' (FBOTs; \citealt{Drout+14,Margutti+19,Ho+21}). New discoveries are likely to continue in the future, with planned surveys by the Large Synoptic Survey Telescope on Vera C. Rubin Observatory \citep{Ivezic+19} and Nancy Grace Roman Space Telescope \citep{Spergel+15}, as well as wide-field UV satellite missions such as ULTRASAT \citep{Sagiv+14}, UVEX \citep{Kulkarni+21}, and QUVIK \citep{Werner+23}. 

A feature common to many transient events is that the progenitor system prior to the explosion can be shrouded in an opaque shell of gas and dust.  In SNe, such a dusty medium can form within the wind of the progenitor star (e.g., \citealt{Chevalier&Fransson94}), whose mass-loss rate may be enhanced in the days to years leading up to the explosion as a result of physical processes driven by late-stage nuclear burning (e.g., \citealt{Quataert&Shiode12,Arnett&Smith14,Matsumoto&Metzger22b}). Alternatively, a small fraction of massive stars may explode into residual accretion disks left over from their birth environments \citep{Metzger10} or from an earlier binary common-envelope phase (e.g., \citealt{Tuna&Metzger23}). In stellar mergers, mass-loss from the donor star leading up to dynamical coalescence event, can enshroud the binary in a weakly-bound or slowly-expanding dusty shell (e.g., \citealt{Pejcha+16a,Pejcha+16b, Pejcha+17}). TDEs occur within the circumnuclear medium surrounding the supermassive black hole, fed by gas from the nuclear star cluster (e.g., \citealt{Generozov+15}). Most TDEs are expected to arise in quiescent galactic nuclei \citep{Stone&Metzger16}, though many will occur in Active Galactic Nuclei (AGN; e.g., \citealt{Kaur&Stone24}).  AGN harbor ``dusty torii'' extending from near the sublimation radius on sub-pc radial scales from the central black hole (e.g., \citealt{Barvainis87}) out to larger scales of several pc (e.g, \citealt{Jaffe+04}).

A dense and compact dusty medium surrounding the progenitor can partially or completely block out its optical/UV emission in the pre-transient (``quiescent'') state, and during the earliest phase of the transient's rise.  During this period, a substantial fraction of the system's luminosity is absorbed and re-emitted at infrared (IR) wavelengths, where the opacity is lower than in the optical/UV. Examples include the IR-luminous dust-shrouded progenitors of ILRTs such as SN 2008S and NGC 300-OT  (\citealt{Thompson+09,Berger+09,Kochanek11}), which may represent explosions or mergers of stars self-obscured by dense, dusty winds (e.g., \citealt{Yoon&Cantiello10}).  In heavily dust-shrouded TDEs such as PS16dtm \citep{Blanchard+17}, the IR light-curve begins rising before the TDE has become optically bright \citep{Jiang+17}.   

As a transient rises towards peak luminosity, the dust present around it is heated and progressively destroyed to larger radii (e.g., \citealt{Waxman&Draine00,Rosalba&Perna02,Jencson+19}). Eventually the hot unobscured emission of the transient can escape, rendering the event visible to optical/UV surveys and often enabling its discovery. However, even if the early dust-attenuated phase is missed in real-time, evidence for the presence of dust can still be preserved in the form of an IR ``dust echo'' (e.g., \citealt{Pearce&Evans84,Dwek85}). The latter describes dust-reprocessed emission which reaches the observer, even after the transient's UV/optical emission has already begun to fade, due to the light-travel delay from the dust's location.

In the nearby universe, dust echoes are seen from many SNe (e.g., \citealt{Graham+83,vanDyk13}; perhaps most famously from SN1987; e.g., \citealt{Crotts88,Rank+88}), as well as some novae (e.g., \citealt{Sokoloski+13}). Dust echoes lasting years have been seen following a handful of TDEs (e.g., \citealt{Komossa+09,Lu+16,vanVelzen+16,Dou+17,Sun+20,Jiang+21,Cao+22,Newsom+23}; see \citealt{vanVelzen+21} for a review), with evidence for the dust being arranged in a torus-like geometry \citep{Jiang+19,Hinkle22}.  In some TDEs the IR emission is seen even though the optical/UV emission is completely attenuated by dust (e.g., \citealt{Mattila+18,Kool+20,Masterson+24}). In the well-studied FBOT AT2018cow, a separate component of slowly-decaying near-IR emission lasting several weeks \citep{Perley+19} was interpreted as a dust echo by \citet{Metzger&Perley23}, even though the transient's UV emission showed no evidence for dust reddening upon discovery near peak light (consistent with the dust being completely destroyed during the earlier rise phase). Properly interpreted, such IR signals offer a probe of the gaseous environments surrounding FBOT sources, providing new insights into their origins.

Several past theoretical studies address the dust-reprocessing signal from transients (e.g., \citealt{Dwek85,Heng+07,Kochanek23}). Most of these works either employ multi-frequency radiation transport simulations under a steady-state assumption \citep{Kochanek23}, or include time-dependent heating and light travel-time effects but assume the absorbing dust shell is optically-thin to the reprocessed radiation (e.g., \citealt{Heng+07}). Neither of these approaches is necessarily ideal for rapidly-evolving transients embedded in (what is, at least initially) a highly-opaque envelope, such as FBOTs, ILRTs, and some TDEs.  

Here we present 2D axisymmetric time-dependent radiation transport simulations with \texttt{Athena++} \citep{Stone+20, Jiang21} of dust-shrouded transients, which follow the heating and sublimation of the gas by a rising central radiation source and calculate the echo signal seen by a distant observer accounting for light travel time effects. Although our calculations are in some ways the first of their kind, we make several simplifying assumptions in this initial work, particularly the adoption of frequency-integrated (grey) opacities. As a result, we are not able to obtain detailed predictions for the color or spectral evolution of the reprocessed radiation, relative to existing specialized dust radiative transfer codes such as DUSTY \citep{IvezicElitzur97}, RADMC-3D \citep{Dullemond+12}, or 2-DUST \citep{Ueta&Meixner03}. Nevertheless, our approach enables an exploration of effects such as the interplay between heating/sublimation and radiation diffusion, and the role of the dust geometry on the reprocessed light-curve shape. 

This paper is organized as follows.  In Sec.~\ref{sec:analytical_estimates} we overview the physical setup of a luminous hot source embedded in a dusty envelope, and estimate the properties of the dust echo signal under different conditions. In Sec.~\ref{sec:methods} we detail our numerical simulations. In Sec.~\ref{sec:results} we present and interpret the results of a suite of numerical models which span the regimes of fast- and slow-rising transients, and for spherical and torus-shaped dust geometries. As an example application, we present model fits to the IR light-curve of AT2018cow (Sec.~\ref{sec:AT2018cow}).  We summarize our conclusions in Sec.~\ref{sec:conclusions}.

\section{Physical Setup}
\label{sec:analytical_estimates}

We first give a semi-analytical description of the expected evolution of a spherically symmetric system, though this assumption will be relaxed in our numerical simulations.  We consider reprocessing by an (initially) optically-thick dust shell situated close to the source. Estimates of the reprocessing signal by optically-thin dust at larger radii, and extending to later times, are described in Appendix~\ref{sec:opticallythin}. The distinct optically-thick and optically-thin light-echo phases are summarized in Fig.~\ref{fig:echo_qualitative}. 

Prior to the rise of the transient, the medium of gas and dust surrounding the explosion is assumed to possess a radial mass density profile of the form,
\begin{align}
    \rho(r) \propto r^{-p}, \quad r > r_{\rm in}
    \label{eq:rhoradial}
\end{align}
exterior to some inner radius $r_{\rm in}$, where $p > 1$. Examples of such a medium in the case of SNe include a steady-wind from the progenitor star $(p = 2)$, or one whose mass-loss rate is rising in time leading up to the explosion ($p > 2$). In TDEs, the surrounding gas would represent an AGN torus or circumnuclear medium (e.g., $p = 3/2$ for Bondi accretion, though larger values of $p \approx 3$ are inferred by modeling TDE radio afterglows; see \citealt{Alexander20}, their Fig.~2). We fiducially take $p = 3$, motivated by TDEs and the circumstellar media surrounding FBOTs on radial scales of light-weeks (e.g., \citealt{Ho+19,Margutti+19}) relevant to a putative dust-echo signal \citep{Metzger&Perley23}.

The gas is initially in equilibrium with radiation emitted as a steady luminosity $L_{\rm Q}$ from the central progenitor in its pre-transient ``quiescent'' state, which thus defines the initial radial temperature profile $T(r)$ of the dust. The optical depth external to a given radius $r$ is given by
\begin{align}
    \tau(r) = \int_{r}^{\infty}\,\kappa(\rho,T)\,\rho\,dr^\prime 
    \label{eq:tau_r}
\end{align}
where $\kappa$ is the opacity.  Because our gray transport calculations adopt the Rosseland mean opacity at the gas temperature, $\tau$ here only represents the optical depth seen by the reprocessed radiation in thermal equilibrium with the gas/dust, not the typically higher optical depth seen by the direct hotter UV radiation from the transient.

Dust is assumed to exist below a critical sublimation temperature, $T_{\rm sub}$, whose value depends on the properties of the dust-forming species (for the dust composition and corresponding opacity law used in most of our simulations, $T_{\rm sub} \approx 10^3\,$K). When dust is present, its opacity $\kappa = \kappa_{\rm d} \approx 10-100$ cm$^{2}$ g$^{-1}$ dominates over other sources of gas opacity (see Fig.~\ref{fig:opacity_plot}).  

The transients we consider typically rise to peak luminosity on a timescale $t_{\rm pk}$ from days to months (an explicit form for the transient's intrinsic UV/optical light-curve $L_{\rm tr}(t)$ will be given below). We are justified in neglecting the effects of radiation pressure from the transient on the dynamics of the gas surrounding the source because such acceleration occurs over a timescale,
\begin{align}
    & t_{\rm{dyn}} \equiv \frac{4\pi r^2 c v}{\kappa L_{\rm tr}} \approx \frac{c v}{\kappa \sigma T_{\rm sub}^{4}}\left(\frac{r}{r_{\rm sub}}\right)^{2} \nonumber \\
    &\approx 2.5\times 10^{7}{\rm s}\,\left(\frac{v}{100\,{\rm km\,s^{-1}}}\right)\left(\frac{\kappa}{100\,\rm cm^{2}\,g^{-1}}\right)^{-1}\left(\frac{T_{\rm sub}}{10^{3}\,\rm K}\right)^{-4}
\end{align}
which is much longer than $t_{\rm pk}$. Here, $v$ is the radial velocity of the gas prior to the transient, $c$ is the speed of light, and in the final line we have normalized radii of interest to the approximate sublimation radius $r_{\rm sub} \approx (L_{\rm tr}/8\pi \sigma T_{\rm sub}^{4})^{1/2}$, as described below. 

On the other hand, dust and gas are heated by the transient light on a very short timescale,
\begin{align}
    t_{\rm{th}} \approx \frac{k_B T}{\kappa m_p a cT^4} \approx 3\,\text{s}\,\left(\frac{\kappa}{100\,\text{cm}^2\,\text{g}^{-1}}\right)^{-1}
    \left(\frac{T}{10^3\,\rm K}\right)^{-3},
\end{align} 
where $a$ is the radiation constant and $m_p$ the proton mass. The radiation and gas temperatures thus remain tightly coupled as the transient brightens. As a result, the dust effectively instantaneously reprocesses the incident UV/optical transient light into IR emission at the gas temperature, until the dust is heated enough to be destroyed and the opacity drops.

The dust present at a given radius $r$ is sublimated once the local energy density exceeds $u_\gamma = aT_{\rm sub}^4$. Estimating half of this energy to be supplied by the incident UV light of the transient, of energy density $u_{\gamma} = L/(4\pi r^{2}c)$, and the other half from reprocessed radiation, we estimate that a critical luminosity $L_{\rm sub} \approx 2\pi r^2 a c T_{\rm sub}^4$ is required for sublimation. An instantaneous ``sublimation radius'' can therefore be defined:
 \begin{align}
    r_{\rm sub}(t) = \left(
    \frac{L_{\rm tr}(t - r_{\rm sub}(t)/c)}{2\pi\,a\,c\,T_{\rm sub}^4}\right)^{1/2}
    \label{eq:rsub_t},
 \end{align}
 i.e., by equating $L_{\rm sub}$ to the transient's emitted luminosity at the retarded time $t - r_{\rm sub}(t)/c$. 
 
 As the transient rises, initially cold dust upstream of the growing sublimation front is gradually heated up and eventually destroyed. Because $\kappa \approx 0$ for $r < r_{\rm sub},$ the optical depth through the dust shell (Eq.~\eqref{eq:tau_r}),
 \begin{align}
    \tau(t) = \int_{r_{\rm sub}(t)}^\infty\,\kappa\,\rho \,dr
    \approx \left(\frac{r_{\rm sub}(t)}{R_{\rm ph,0}}\right)^{1-p},
    \label{eq:tau_inf}
 \end{align}
decreases in time, where here $R_{\rm ph,0}$ is the initial (pre-transient) photosphere radius, defined according to:
 \begin{align}
    \tau\left(R_{\rm ph,0}\right) = 1.
    \label{eq:rph_0}
 \end{align}
 The dust becomes optically-thin to the reprocessed radiation once the transient reaches a critical luminosity
 \begin{align}
     L_{\rm tr} &\gtrsim L_{\rm thin} \equiv 8\,\pi\,R_{\rm ph,0}^2\,\sigma T_{\rm sub}^4 \nonumber \\
     &\approx {1.4\times 10^{43}\,\rm erg\,s^{-1}}\left(\frac{T_{\rm sub}}{10^{3}\,{\rm K}}\right)^{4}\left(\frac{R_{\rm ph,0}}{10^{17}\,\rm cm}\right)^{2},
     \label{eq:L_thin}
 \end{align}
Note that $R_{\rm ph,0}$ and hence $L_{\rm thin}$ are mainly determined by the properties of the environment prior to the transient, particularly the radial density profile and dust properties.

Most of the total energy reprocessed by an optically-thick dust shell occurs when the transient is at its highest luminosity just prior to sublimating the photosphere, i.e. when $L_{\rm tr} \sim L_{\rm thin}$. Adopting a common transient rise-law of the form
 \be
L_{\rm tr}(t) = L_{\rm pk}\left(\frac{t}{t_{\rm pk}}\right)^{2}, \,\,\, t < t_{\rm pk},
\label{eq:LUVrise}
\ee
the transient spends a duration
\be
t_{\rm rise} \approx \frac{t}{2} \approx \left(\frac{L_{\rm thin}}{L_{\rm pk}}\right)^{1/2}t_{\rm pk}
\label{eq:t_rise}
\ee
at luminosities $L_{\rm tr} \sim L_{\rm thin}.$  This ``rise-time'' near the photosphere sublimation luminosity is a critical parameter, as it sets the timescale over which the majority of the reprocessed energy is created, at radii $r_{\rm sub} \approx R_{\rm ph,0}$.

From Eqs.~\eqref{eq:rsub_t} and \eqref{eq:LUVrise}, the sublimation front $r_{\rm sub}$ is found to expand at a constant speed:
\begin{align}
    \frac{ v_{\rm sub}}{c} \equiv \frac{r_{\rm sub}}{ct} = \frac{t_{\rm lc}}
    {t_{\rm lc} + 2t_{\rm rise}}
    \label{eq:vsub}
\end{align}
where 
\begin{align}
        t_{\rm lc} \equiv \frac{R_{\rm ph,0}}{c}
        \label{eq:t_lc}
\end{align}
is the light crossing-time of the photosphere.

\begin{figure}
    \centering
    \includegraphics[width=0.5\textwidth]{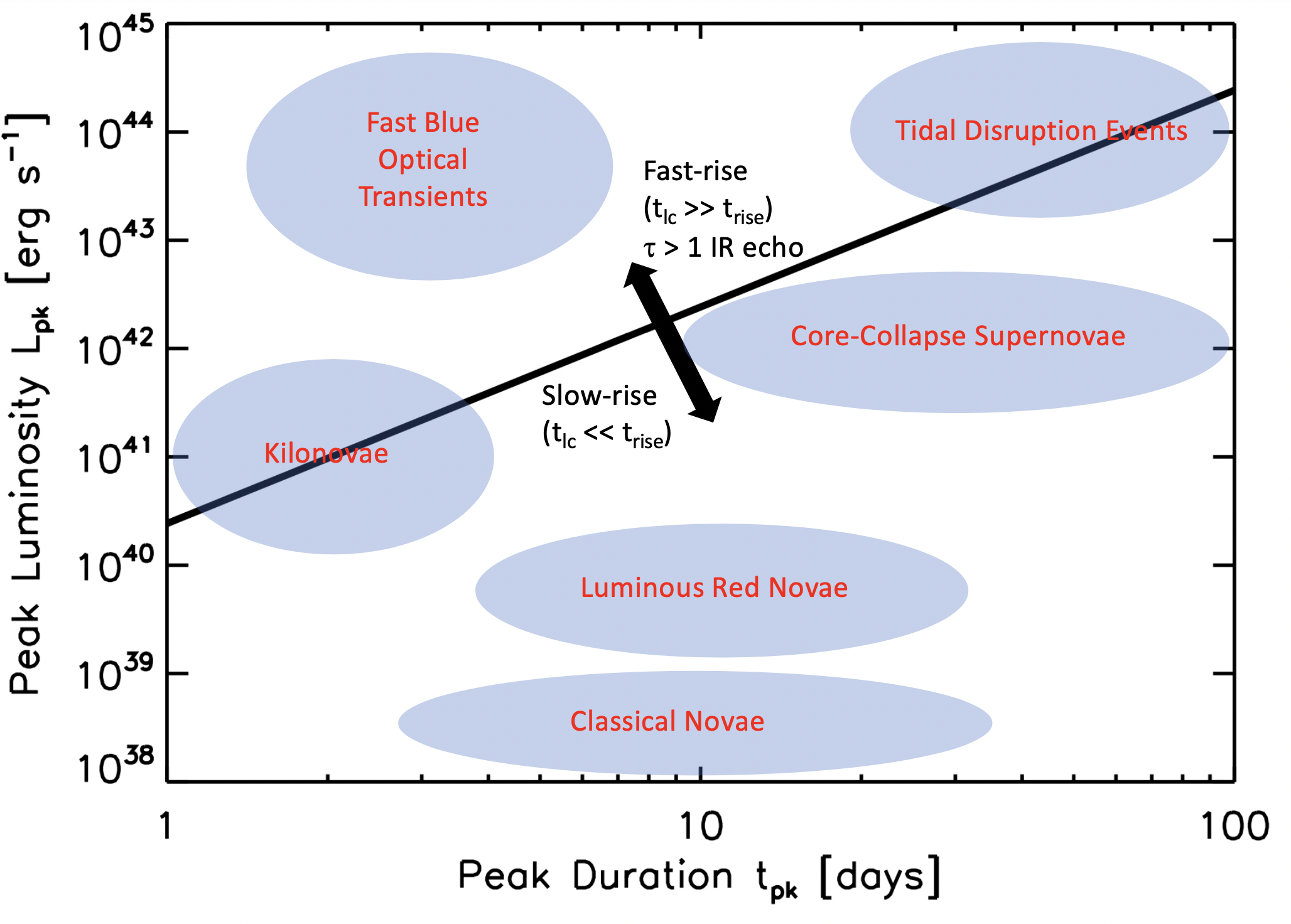}
    \caption{Classes of optical/UV transients in the space of peak luminosity and duration. A black curve denotes the boundary $t_{\rm lc} = t_{\rm rise}$ (Eq.~\eqref{eq:trise_over_tlc}, for $T_{\rm sub} = 1500$ K) separating events for which the reprocessed emission from a optically-thick dust shell surrounding the progenitor will out the transient itself as an ``echo'' ($t_{\rm lc} \gg t_{\rm rise}$) versus cases in which light-crossing delays are negligible ($t_{\rm lc} \ll t_{\rm rise}$) and hence the IR light-curve will more closely track the transient light-curve prior to dust destruction (Fig.~\ref{fig:echo_qualitative}).}
    \label{fig:transients_parameterspace}
\end{figure}

The appearance of the transient and its reprocessed radiation to an external observer will exhibit different characteristics, depending on whether the transient is rising rapidly ($t_{\rm rise} \ll t_{\rm lc}$)
or slowly ($t_{\rm rise} \gg t_{\rm lc}$) compared to the light-crossing time of the photosphere. This critical ratio can be expressed entirely in terms of the transient and dust properties according to:
\begin{align}
\frac{t_{\rm rise}}{t_{\rm lc}} = \frac{ct_{\rm pk}}{R_{\rm ph,0}}\left(\frac{L_{\rm thin}}{L_{\rm pk}}\right)^{1/2} = \left(\frac{4\pi (c t_{\rm pk})^{2}\sigma T_{\rm sub}^{4}}{L_{\rm pk}}\right)^{1/2}.
\label{eq:trise_over_tlc} 
\end{align}
The location of different transient classes in the space of $t_{\rm pk}, L_{\rm pk}$ and the location of the $t_{\rm rise} = t_{\rm lc}$ boundary for $T_{\rm sub} = 1500$ K are shown in Fig.~\ref{fig:transients_parameterspace}.  We see that both FBOTs and TDEs can reside in the fast-rise regime ($t_{\rm rise} \ll t_{\rm lc}$), while LRN and classical nova typically reside in the slow-rise regime ($t_{\rm rise} \gg t_{\rm lc}$). We now discuss the observable properties of the reprocessed radiation separately in these two limits.

\begin{deluxetable}{cccc}
\tablecolumns{3}
\tablewidth{0pt}
 \tablecaption{Time- and luminosity-scales of reprocessed radiation in the fast-rise (Sec.~\ref{sec:fast_rise}) and slow-rise (Sec.~\ref{sec:slow_rise}) limits.
 \label{tab:timescales_luminosities}}
 \tablehead{
 \colhead{Symbol} & \colhead{Description} & \colhead{Fast ($t_{\rm rise} \ll t_{\rm lc}$)} & \colhead{Slow ($t_{\rm rise} \gg t_{\rm lc}$)}}
 \startdata 
 $t_{\rm on}$ & Onset Time & $t_{\rm lc}$ (Eq.~\eqref{eq:t_lc}) & $t_{\rm esc}$ (Eq.~\eqref{eq:t_esc}) \\
 $t_{\rm IR}$ & Duration & 2$t_{\rm lc}$ & $t_{\rm rise}$ (Eq.~\eqref{eq:t_rise}) \\
 $L_{\rm IR}$ & Luminosity & $L_{\rm thin}(t_{\rm rise}/t_{\rm IR})$ & $L_{\rm thin}$ \\
 \enddata
\end{deluxetable}

\subsection{Fast-Rise Regime ($t_{\rm rise} \ll t_{\rm lc}$)}
\label{sec:fast_rise}

When the transient rises rapidly, the sublimation front moves close to the speed of light 
$v_{\rm sub} \approx c$. The energy reprocessed on a given radial scale $\sim r_{\rm sub}$ or time scale $r_{\rm sub}/v_{\rm sub} \sim t$ (e.g., as the sublimation front doubles from $r_{\rm sub}(t)$ to $\sim 2 r_{\rm sub}(t)$), grows as:
\begin{align}
    E_{\rm rep}(t) \sim L_{\rm tr}(t)\,t \propto t^3.
    \label{eq:en_rep}
\end{align}
In the fast-rise limit, the reprocessed radiation energy essentially free-streams in all directions, since the sublimation front clears out upstream
opacity moving at roughly the same speed as the reprocessed photons. Since no photons get out until the photosphere is sublimated, the light-curves of both the escaping UV transient emission, as well as the reprocessed IR emission, only begin to first rise above the quiescent luminosity on a timescale $t_{\rm on} \approx t_{\rm lc} \gg t_{\rm rise}$ defined by the finite light-crossing time of the reprocessing region $\sim R_{\rm ph,0}$. 

The reprocessed radiation received by an external observer receives contributions from each radial dust shell at its respective sublimation luminosity, accounting for relative delays in their arrival times to the observer. As Eq.~\eqref{eq:en_rep} reveals, most of the reprocessed energy occurs at the largest radii ($r \sim R_{\rm ph,0}$) where $L_{\rm tr} \sim L_{\rm thin}$ (Eq.~\eqref{eq:L_thin}).  The total absorbed energy can thus be approximated as
\begin{align}
    E_{\rm rep}(t_{\rm rise}) 
    \approx L_{\rm thin}\,t_{\rm rise}.
\label{eq:Erepmax}
\end{align}
In the rapid-rise limit, this energy reaches a distant observer over a timescale corresponding to the light-travel time delay $2t_{\rm lc} \gg t_{\rm rise}$ between the farthest and nearest portions of the photosphere. The IR echo therefore exhibits a characteristic duration $t_{\rm IR} \approx 2t_{\rm lc}$ and luminosity, 
\be L_{\rm IR} \approx \frac{E_{\rm rep}(t_{\rm rise})}{2t_{\rm lc}} \approx L_{\rm thin}\frac{t_{\rm rise}}{2t_{\rm lc}} \approx \frac{L_{\rm thin}^{3/2}}{L_{\rm pk}^{1/2}}\frac{t_{\rm pk}}{t_{\rm IR}}
\label{eq:LIR_fast}.
\ee

\subsection{Slow-Rise Regime ($t_{\rm rise} \gg t_{\rm lc}$)}
\label{sec:slow_rise}

For slowly-rising transients ($t_{\rm rise} \gg t_{\rm lc}$), reprocessing occurs over a timescale longer than the light-travel time delay to the observer. Hence, the IR light-curve is expected to track the transient light-curve and peak on the same timescale $t_{\rm IR} \approx t_{\rm rise}$ the photosphere is sublimated, and at a luminosity $L_{\rm IR} \approx L_{\rm thin}$ which is comparable to that of the transient around this same time.

However, the reprocessed IR light-curve will not in general track that of the transient during the earliest phases after the explosion, at times $\ll t_{\rm rise}$.  As a result of the finite photon diffusion-time through the dust shell, the IR light-curve will only first begin to rise above the quiescent luminosity with some delay, $t_{\rm on}$, after the onset of the transient. In the slow-rise regime ($t_{\rm rise} \gg t_{\rm lc}$), the sublimation front propagates well below the speed of light,
$v_{\rm sub} \approx (t_{\rm lc}/t_{\rm rise})\,c$ (Eq.~\eqref{eq:vsub}). 
This gives the radiation being absorbed and reprocessed at $\approx r_{\rm sub}$ a chance to diffuse ahead of the sublimation front, escaping the photosphere before it is destroyed, rendering direct light from the transient visible. While the dust ahead of the sublimation front is still optically-thick ($\tau > 1$), photons diffuse outwards at a velocity $v_{\rm diff}(t) \equiv c/\tau(t)$. By equating $v_{\rm diff} \approx v_{\rm sub}(t)$, and using Eq.~\eqref{eq:tau_inf} for $\tau(t)$, we see that photons can first outrun the sublimation front, thus reaching the external observer, after a critical escape time:
\begin{align}
    t_{\rm esc} &\equiv 
    t_{\rm lc}
    \left(\frac{c}{v_{\rm sub}}\right)^{\frac{(p-2)}{(p-1)}} = t_{\rm lc}^{1 - \frac{(p-2)}{(p-1)}}
    \left(t_{\rm lc} + 2t_{\rm rise}\right)^{\frac{(p-2)}{(p-1)}}
    \nonumber\\
    &\underset{p=3}\approx \left(2t_{\rm lc}t_{\rm rise}\right)^{1/2} \approx \left(\frac{t_{\rm lc}}{2t_{\rm rise}}\right)^{1/2}\,2t_{\rm rise}.
    \label{eq:t_esc}
\end{align}

\begin{figure*}
    \centering
    \includegraphics[width=1.0\textwidth]{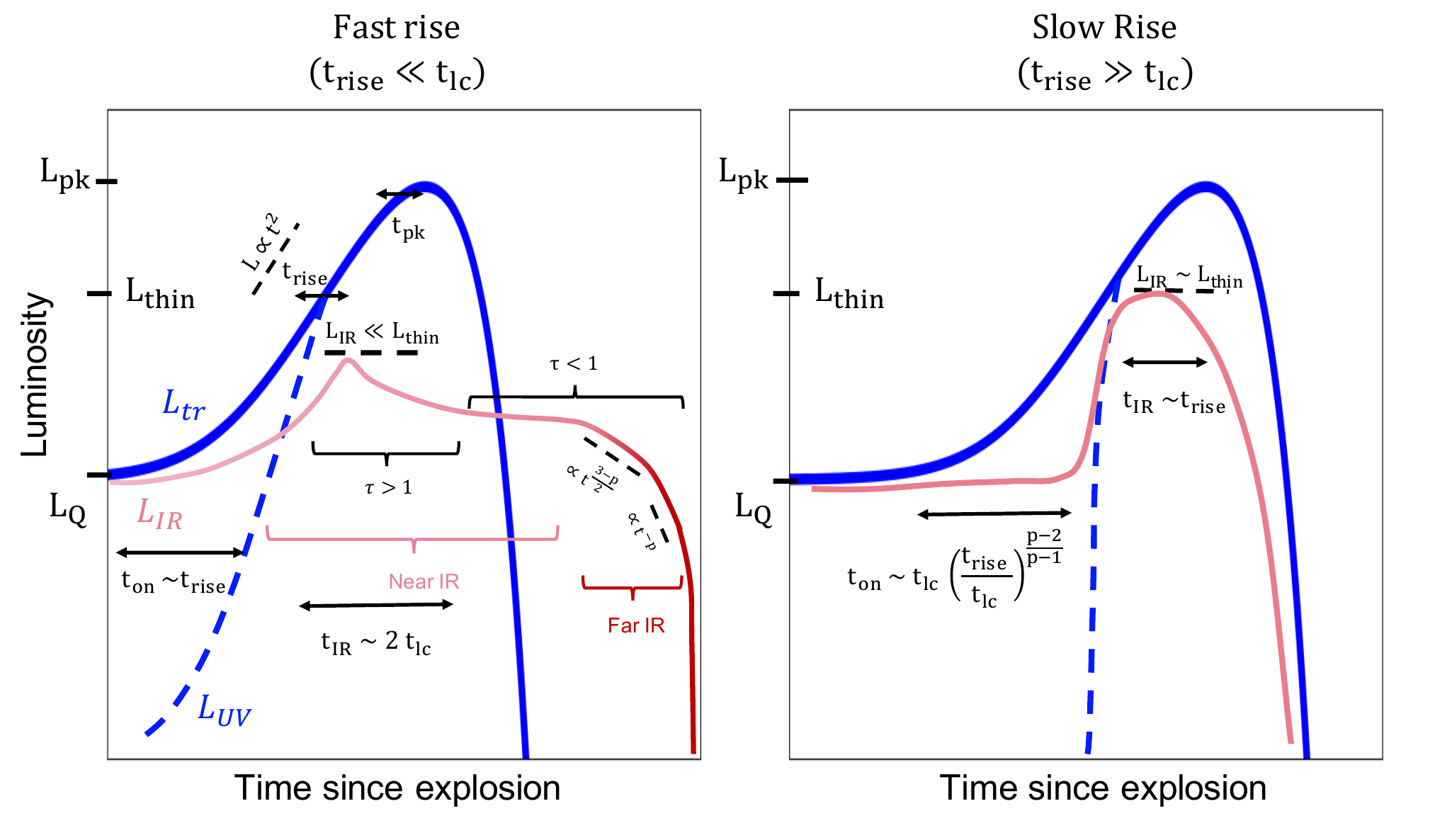}
    \caption{Schematic illustration of the intrinsic transient luminosity $L_{\rm tr}(t)$ (solid blue line), escaping optical/UV light-curve $L_{\rm UV}(t)$ (dashed blue line), and dust-reprocessed IR light-curve $L_{\rm IR}(t)$ for an assumed dust density profile $\rho \propto r^{-p}$.  For $L_{\rm IR}$, pink and red colors represent NIR and far-IR emission from hot dust close to the sublimation temperature and cooler dust, respectively. The separate panels depicts the two regimes which control the light-curve shape, depending on the ratio of the transient rise-time $t_{\rm rise}$ near the sublimation luminosity $L_{\rm thin}$ (Eq.~\eqref{eq:L_thin}) to the light crossing-time of the (initial) dust photosphere $t_{\rm lc}$ (Eq.~\eqref{eq:trise_over_tlc}; see Sec.~\ref{sec:analytical_estimates}). The transient $L_{\rm tr}(t)$ rises from the quiescent luminosity $L_{\rm Q}$ up to $L_{\rm pk}$ over a timescale $t_{\rm pk}$. The reprocessed IR light-curve first rises on a timescale $t_{\rm on}$, which in the slow-rise case is delayed from the onset of the explosion due to photon diffusion effects. After IR emission from the initially optically-thick dust shell reaches the observer, radiation from the optically-thin dust which extends above the original dust photosphere continues the echo signal to later times and, ultimately, longer IR wavelengths (the portion of the light-curve marked $\tau < 1$; Appendix \ref{sec:opticallythin}).}
    \label{fig:echo_qualitative}
\end{figure*}

In summary, in the fast-rise regime the reprocessed emission rises to peak on the same timescale $t_{\rm rise}$ the photosphere is sublimated and the transient UV/optical light first escapes without dust attenuation.  By contrast, in the slow-rise case the reprocessed radiation starts rising earlier, i.e. $t_{\rm on} \approx t_{\rm esc} \lesssim t_{\rm IR} \approx 2\,t_{\rm rise}$ (though still delayed relative to the explosion/onset of the transient emission). The light-curve evolution in the ``fast-rise'' and ``slow-rise'' limits, and their key time- and luminosity-scales are 
summarized in Fig.~\ref{fig:echo_qualitative} and Table~\ref{tab:timescales_luminosities}, respectively.

\begin{deluxetable*}{cccc}
\tablecolumns{3}
\tablewidth{0pt}
 \tablecaption{Summary of model parameters and their values in the fiducial model $\texttt{SPHERE$\_$FAST}$ \label{tab:parameters_symbols}}
 \tablehead{
 \colhead{Symbol} & \colhead{Parameter} & \colhead{Equation} & \colhead{$\texttt{SPHERE$\_$FAST}$}}
 \startdata 
  $\rho_0 = \rho(r_{\rm in})$  & density normalization of external medium & \eqref{eq:rhoangle} & $10^{-16}\,\text{g}\,\text{cm}^{-3}$\\
  $p$  & radial power-law index of external medium density profile  & \eqref{eq:rhoangle} & $3$\\
  $r_{\rm in}$  & inner radius of external medium & \eqref{eq:rhoangle} & $10^{16}\,
  \text{cm}$\\
     $\Delta$ & angular extent of external medium & \eqref{eq:rhoangle} & $\infty$ \\
    $X_d$ & dust-to-gas mass ratio & - & $0.1$ \\
    $a_{\rm max}$ & maximum dust grain size & \eqref{eq:dust_size} & $1\mu m$ \\
  $L_{\rm Q}$ & quiescent luminosity prior to transient & \eqref{eq:lum_bdry_cond} & $10^{39}\,\text{erg}\,\text{s}^{-1}$ \\
 $L_{\rm pk}$ & peak luminosity of transient & \eqref{eq:lum_bdry_cond} & $4\,\times\,10^{42}\,\text{erg}\,\text{s}^{-1}$ \\
   $t_{\rm pk}$  & peak timescale of transient & \eqref{eq:lum_bdry_cond} & $7.1\,\text{day}$ \\
   \hline 
&   Derived Parameters & \\
   \hline
   $M$ & total mass of dust-laden external medium & - & 
   $0.15\,M_\odot$ \\
 $R_{\rm ph,0}$ & initial photosphere radius of dust & \eqref{eq:rph_0} & $2.6 \times 10^{16}\,\text{cm}$ \\
 $T_{\rm sub}$ & sublimation temperature (of limiting grain species, olivine) &  - & $1000 K$ \\
 $L_{\rm thin}$ &  transient luminosity at which dust is sublimated at photosphere $R_{\rm ph,0}$ & \eqref{eq:L_thin} & $9.6 \times 10^{41}\,\text{erg}\,\text{s}^{-1}$ \\
 $t_{\rm rise}$ & transient rise-time when photosphere sublimates, i.e. when $L_{\rm tr} \approx L_{\rm thin}$ & \eqref{eq:t_rise} & $1.9\,\text{days}$ \\ 
 $t_{\rm on}$ & delay after explosion before initial escape of reprocessed IR radiation & \eqref{eq:t_esc} & $\approx 10\,\text{days}$ \\ 
 $t_{\rm lc} \equiv R_{\rm ph,0}/c$ & light-crossing time of initial dust photosphere & \eqref{eq:t_lc} & $10\,\text{days}$
 \enddata
\end{deluxetable*}

\section{Methods}
\label{sec:methods}

We use the implicit radiation transport module of \texttt{Athena++} \citep{Jiang21} to simulate dust reprocessing by a central transient source. \texttt{Athena++} solves the equations of hydrodynamics coupled to equations of radiative transfer.  We refer the reader to \cite{Jiang21} for details of the code, and describe our particular
setup in the following.

\subsection{Spatial domain and resolution}
 \label{sec:resolution}
Our models are axisymmetric, hence we solve the equations in two 
dimensions, using spherical polar coordinates. We shall adopt a steep radial density profile for the external medium, such that the total mass and optical depth are concentrated at small radii $r = r_{\rm in}$ (the density distribution is introduced in Eq.~\eqref{eq:rhoangle} below). Our computational domain consists of a logarithmically spaced grid of $n_r = 1200$ grid points extending from $0.3 \, r_{\rm in}$ to 
$10 \, r_{\rm in}$. We use $n_\theta = 32$ grid points
uniformly discretized in the $\theta$ direction.

For discretization of the ray directions, we use an angular grid labeled by the coordinates $(\zeta, \psi)$ defined locally in
terms of the spherical coordinates $(r, \theta, \phi)$ \citep{Davis&Gammie20, Jiang21}
\begin{align}
    \cos\zeta = \vu{r}\cdot\vu{n}\\
    \cos\psi = \vu{\theta} \cdot \vu{n}
\end{align}
We use $n_\zeta = 40$ and $n_\psi = 4$ rays in a quadrant of the unit sphere, leading
to a total of 640 angles spanning all directions about a given point $(r, \theta, \phi)$.
We customized this
angular grid by adding two more directions corresponding to angles $\vu{n} \cdot \vu{r} = \pm 1$ with appropriate weights, where $\mathbf{r}$ is the position unit vector and $\vu{n}$ is the ray direction. 

The custom angular grid described above was necessary in our calculations
for two reasons: (i) The intensity field at the inner boundary represents
light incident from a far-away source, hence it is concentrated
in the radial direction $\vu{n} \cdot \vu{r} = 1$ 
(see Sec.~\ref{sec:initial_and_bdry_cond}). Accurate representation of this boundary condition would require a prohibitively high angular resolution. (ii) In the absence of scattering effects, purely radial ($\cos \zeta = 0$) rays travel along straight paths on a discrete, spherical polar grid, whereas any non-radial ray ($\cos \zeta \neq 0$) travels in zig-zag paths, leading to numerical leakage of the radiation energy. This prevents accurate extraction
of the dust-reprocessed energy from the simulation output in a frequency-integrated calculation (Sec.~\ref{sec:opacity}), since the intensity is dominated by the (in reality) optical/UV radiation of the transient
peaking at $L_{\rm pk} \gg L_{\rm thin}$, which is distributed at all frequencies (including the IR band) according to Planck function when grey opacities are used. On the other hand, the custom grid including purely radial rays allows us to distinguish between the source (``UV'') and reprocessed (``IR'') energies by ray direction: To a good approximation, purely radial rays represent UV rays, and all non-radial rays represent the reprocessed IR radiation. Therefore, we use (outgoing) radial rays when computing UV light-curves, and non-radial rays for reprocessed IR light-curves (see Sec.~\ref{sec:light_curve_computation}).

\subsection{Initial and boundary conditions} 
\label{sec:initial_and_bdry_cond}

We consider a central source with a finite (quiescent) luminosity $L_{\rm Q}$ at $t = 0$, which then rises as the transient brightens like $L(t) \sim (t/t_{\rm rise})^2$ at times around the critical sublimation luminosity $L \sim L_{\rm thin}$ (Eq.~\eqref{eq:LUVrise}), before peaking at $L_{\rm pk}$ on a timescale $t_{\rm pk}$ and then subsequently decaying away exponentially at later times. We smoothly interpolate between these limits using the functional form:
\begin{align}
    L_{\rm tr}(t) = L_{\rm Q}\left(1 + \frac{t}{t_0}\right)^2
    \exp\left(-t/t_1\right). \label{eq:lum_bdry_cond}
\end{align}
 We choose the set 
 $\{L_{\rm Q}, L_{\rm pk}, t_{\rm rise}\}$ 
 as the control parameters for the transient evolution, and set the values
 of $\{t_0, t_1\}$ accordingly.  

In the transient classes we consider (SNe, TDE, FBOT), the central luminosity source typically originates from an expanding shell of ejecta located at scales much smaller than the radial extent of the dust-forming region we model in our simulations. Hence, the incident rays are strongly focused in the radial direction. We impose the following boundary condition on the intensity field at the inner boundary of our domain: We let the rays leaving the grid ($\cos\zeta < 0$) free-stream, by filling out the ghost cell values of rays in these directions with their values in the innermost active cell. Then, we compute the flux leaving the grid, and set the values of the radial rays entering the grid to be such that the net flux gives the desired luminosity in Eq.~\eqref{eq:lum_bdry_cond}. We set the values of
all other rays $0 < \cos\zeta < 1$ to zero. 

At the outer boundary of the computational domain, we set the incoming 
$(\cos\zeta < 0)$ rays to zero, and copy outgoing
intensities from the last active cell to the ghost zones.

We start our simulations with a static ($\mathbf{v} = 0$) gas
of mass density distribution in spherical polar coordinates $(r, \theta)$
\begin{align}
    &\rho(r, \theta) = \rho_0
     \left(\frac{r}{r_{\rm in}}\right)^{-p}
     \exp\left(-\left(\frac{r_{\rm in}}{r}\right)^2\right)\nonumber\\
    &\quad \times \frac{\exp\left(-\left(\frac{\cos\theta}{2\Delta}\right)^2\right)}{2\int_0^1\,d\mu\,\exp(-\mu/(2\Delta^2))}
    \label{eq:rhoangle}
\end{align}
where the free parameters 
are $\{r_{\rm in}, p, \rho_0, \Delta\}$ (see Table \ref{tab:parameters_symbols}). 
This distribution ensures that mass is distributed
according to $\rho \propto r^{-p}$ at $r > r_{\rm in}$,
and concentrated in a solid angle $\Omega \sim 2\pi \Delta$ around the equator $(\theta = \pi/2)$; $\Delta \rightarrow \infty$ corresponds to a spherically symmetric shell.

The simulations are performed in two steps: First, we initialize the intensity field 
to zero, and the gas temperature to be spatially uniform at $T = 100$ K.  We then evolve the system long enough to reach a steady state 
at the (constant) quiescent luminosity, $L = L_{\rm Q}$, imposed at the inner boundary on the intensity field. The gas properties ($\rho$, $T$) and intensity field achieved in this steady-state are then taken as initial conditions on the subsequent transient phase, during which the time-evolving intensity Eq.~\eqref{eq:lum_bdry_cond} is imposed at the inner boundary.

\subsection{Opacity table} \label{sec:opacity}

We employ frequency-integrated (grey) opacities, leaving more accurate multi-band calculations to future work. For the transient systems we consider, the external medium usually represents a slowly-expanding outflow from the progenitor star, or in the TDE case, potentially a slow inwards accretion flow. In either case, the timescale for dust formation prior to the transient is much shorter than the dynamical inflow/outflow time at the radii of solid condensation, such that grain growth up to sizes $a \gtrsim 0.1-1\mu m$ may be possible (e.g., \citealt{Metzger&Perley23}). As described below, we employ tabulated dust opacities corresponding to relatively large assumed grain sizes. On the other hand, the transient decays over a much shorter timescale, making dust {\it reformation} as the transient fades unlikely (however, see \citealt{Kochanek11} for a counterexample). Thus we assume that dust does not reform when the gas cools back down below the dust sublimation temperature. This is achieved by tracking the destruction of a dust species in a cell, and setting the contribution of that species to the opacity to zero in the remainder of the simulation at that cell, if the temperature at any previous time has exceeded the sublimation temperature of that dust species.

\begin{figure}
    \centering
    \includegraphics[width= 0.48\textwidth]{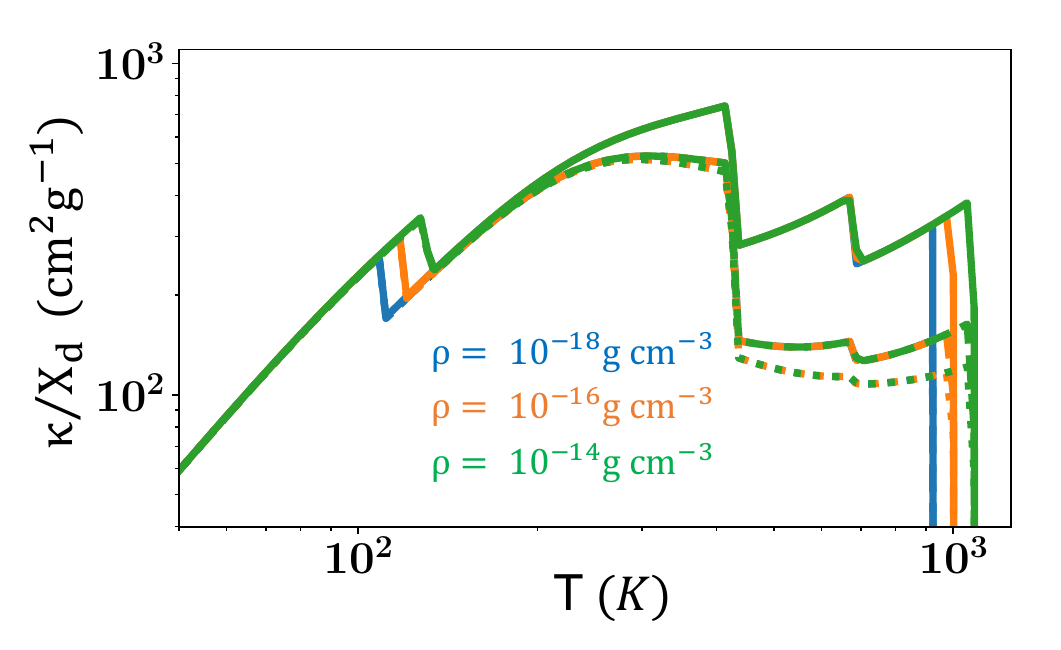}
    \caption{Rosseland-mean of the extinction
    opacity at densities 
    $\rho= 10^{-18}\,\text{g}\,\text{cm}^{-3}$
    (blue),
    $\rho= 10^{-16}\,\text{g}\,\text{cm}^{-3}$
    (orange), 
    $\rho= 10^{-14}\,\text{g}\,\text{cm}^{-3}$
    (green), 
    for assumed maximum grain sizes $a_{\rm{max}} = 10^{-5}\, \text{cm}$ (dotted),
    $a_{\rm{max}} = 3 \times 10^{-5}\, \text{cm}$ (dashed),
    and  $a_{\rm{max}} = 10^{-4}\, \text{cm}$ (solid). The opacity values are normalized to a dust mass fraction $X_d = 1$, but we use $X_d = 0.1$ in our simulations. The opacity drops off sharply above $T_{\rm sub} \approx 10^{3}$ K corresponding to the dust constituent with the highest sublimation temperature.}
    \label{fig:opacity_plot}
\end{figure}

We employ dust opacities as tabulated by the Disk Substructures at High Angular Resolution Project (DSHARP; \citealt{Birnstiel+18}). We now briefly describe the assumptions made in these calculations, referring the reader to the reference for details.
We assume four species of dust grains: water ice, silicates (olivine), troilites, and refractory organic material. For each species,
the absorption and scattering cross sections are
calculated using Mie theory \citep{Bohren&Hoffman98}, for a given grain
size and wavelength. The number $N$ of grains per unit dust size $a$ in the medium is assumed to satisfy the standard power-law shape,
\begin{align}
    \frac{dN}{da} \propto 
    \left(\frac{a}{a_{\rm{min}}}\right)^{-q},
    \quad a_{\rm{min}} \leq a \leq a_{\rm{max}}\label{eq:dust_size}, 
\end{align}
where $a_{\rm{min}} = 0.1\mu m$. We fiducially assume
$q = 3.5$ and $a_{\rm{max}} = 1\mu m$, the latter motivated by estimated maximum grain size in the context of FBOTs
\citep{Metzger&Perley23} or supernova precursor eruptions \citep{Kochanek11}.

The total (wavelength-dependent) opacity at a given gas temperature and mass density is
calculated by summing the contributions of those dust species present (i.e., not sublimated) under these conditions. The density-dependent sublimation temperatures for water ice and olivine are taken from \citet[their Table 3]{Pollack+94}.
The sublimation temperatures of these species tabulated in 
\citet{Pollack+94} extend down to 
$\rho = 10^{-18}\,\text{g}\,\text{cm}^{-3}$. 
At densities below
this value, we use the sublimation temperature at 
$\rho= 10^{-18}\,\text{g}\,\text{cm}^{-3}$ ($T_{\rm sub} = 929\,$K).
Sublimation temperatures for troilite and refractory organics
are taken as $T_{\rm sub} = 680 $K and $425\,$K, respectively \citep{Zhu+21}.

We assume a dust-to-gas mass ratio of $X_d = 0.1$, corresponding to $\approx 10$ times solar metallicity. While this can be motivated in TDEs by the super-solar metallicities of AGN, or in FBOTs by their likely hydrogen-depleted environments (e.g., \citealt{Andrews&Smith18}), $X_{\rm d}$ is largely degenerate with the total density and mass of the dust shell. Fig.~\ref{fig:opacity_plot} shows the temperature dependence of the Rosseland-averaged extinction opacity defined as in \citet[their Eq. 10]{Birnstiel+18} at a few representative densities, for different assumed maximum grain sizes. 

\subsection{Opacity cut-off}
\label{sec:opacity_cutoff}

We focus in this paper on calculating the IR emission dust that is (at least initially) optically-thick. Moving to radii outside the sublimation surface, our use of Rosseland-mean opacities becomes a poorer approximation because the true frequencies of transient's optical/UV light even more greatly exceeds the local gas temperature; even for relatively large dust grains, the dust opacity used in our simulations can be underestimated by up to an order of magnitude for cold $T < 100 $K dust residing at large radii beyond the photosphere. For this reason, we artificially truncate the assumed opacity slightly outside the initial photosphere radius. This restricts the reprocessed radiation we measure to originate from optically-thick regions, where our gray opacity treatment is better motivated. In practice, this is achieved as follows: we initialize our simulation data with the density distribution and steady quiescent luminosity as described in Sec.~\ref{sec:initial_and_bdry_cond}. After evolving the system to steady-state, we locate the initial photosphere radius. We then set the value of opacity to zero at all grid cells beyond this radius throughout the remainder of the simulation. Before the transient rise phase, we evolve the system a little longer so that the new steady state with the changed opacity is reached. This moves the initial photosphere radius slightly inward, consistent with the assumed density distribution in Eq.~\eqref{eq:rhoangle}. We refer to this photosphere radius obtained after the opacity cutoff has been imposed as $R_{\rm ph,0}$.

\subsection{Light-curve calculation}
\label{sec:light_curve_computation}

Construction of the light-curve as seen by a distant observer requires post-processing of the simulation data, since the observer is practically located at infinity, well outside our grid.

To compute the light-curve, we first take a circle of
radius $r \approx R_{\rm ph,0}$ (see Table~\ref{tab:parameters_symbols}) centered at the origin of our 2D grid and construct a 3D sphere by
rotating this circle around
the polar axis. More precisely, the radius is chosen such that the external optical depth satisfies $\tau < 10^{-3}$ at all times and all latitudes. This typically corresponds to a radius outside the initial photosphere radius by $\approx 20\,\%$, for the density distributions 
we consider (Eq.~\eqref{eq:rhoangle}).
The sphere is partitioned into rectangular area elements,
each assigned an intensity distribution from the simulation output. 
Since rays from this surface to the observer do not traverse significant optical depth, they essentially propagate freely towards the observer, hence are conserved along their paths.

Next, we compute the net flux of energy through the hemisphere visible to the observer, at a given observation
time. Since the light travel time to the observer is effectively infinite, we define $t_{\rm obs} = 0$ as the arrival time of the first photon emitted at the inner boundary at $t = 0$ would reach the observer on a direct path (i.e., assuming it were not attenuated). If the observer is located along the direction
$\vu{r}_{\rm obs}$ from the origin, only rays pointing in this direction arrive at the observer, to leading order. A ray emitted in the direction $\vu{n} = \vu{r}_{\rm obs}$ at $\va{r}$ on the sphere and arriving at the observer at $t_{\rm obs}$ was emitted at simulation time $t$ given by
\begin{align}
    t = 
    t_{\rm obs} + \frac{2r}{c}- \frac{\vu{n}\va{r}}{c}
    \label{eq:t_obs}
\end{align}
to leading order in $r/r_{\rm obs}$. For example, consider the emission point $\vu{r} = \vu{n}$: The ray arriving at $t_{\rm obs} = 0$ was emitted from the
sphere at $t = r/c$, which was emitted at the inner
boundary at $t = 0$. 

The contribution from an area element to the net flux satisfies
\begin{align}
    r_{\rm obs}^2\,
    &I_{\rm obs}(t_{\rm obs}, \mathbf{r}_{\rm obs}; \vu{n})
    \,d\Omega_{\rm obs} = \nonumber\\
    &I_{\rm em}(t_{\rm em}, \mathbf{r_{\rm em}}; \vu{n})\,
    \,\vu{n} \cdot 
    d\mathbf{A},\nonumber\\
    &\quad\quad \mathbf{r}_{\rm obs} 
    \cdot d\mathbf{A} > 0
\end{align}
where $t_{\rm em (\rm obs)}, \mathbf{r}_{\rm em (\rm obs)}, I_{\rm em (\rm obs)}$ are emission (observation) time, position and intensity;  $\vu{n}$ is the ray direction, $d\mathbf{A}$ is the area element pointing in the direction of its normal vector,
and $d\Omega_{\rm obs}$ is the solid angle subtended by this area at the observation point. Given $t_{\rm obs}, \vu{r}_{\rm obs}$, we 
determine the value of $I_{\rm em}$, by 
interpolating the simulation output in time,
ray direction and position at the center of
the area element. We store the values of 
the intensity, and projected area separately.
The net flux is found by summing over all
contributions at a given observation time.

\section{Results}
\label{sec:results}

We present results for three simulations, which vary the transient's rise-time and the geometry of the dust shell. The fiducial model \texttt{SPHERE$\_$FAST} parameters are summarized in Table~\ref{tab:parameters_symbols}, while those parameter varied in the models \texttt{SPHERE$\_$SLOW} and \texttt{TORUS$\_$FAST} are summarized in Table~\ref{tab:sim_param}. 

\begin{deluxetable}{ccccc}
\tablecolumns{6}
\tablewidth{0pt}
 \tablecaption{Simulation parameters \label{tab:sim_param}}
 \tablehead{
 \colhead{Run/Parameter}
 & \colhead{$t_{\rm lc}$ (days)}
 & \colhead{$t_{\rm rise}$ (days)}
 & \colhead{$\Delta$}
 & \colhead{$T_{\rm sub}$ (K)}}
 \startdata 
  \texttt{SPHERE$\_$FAST}  & 10 & 1.9 & $\infty$ & 1000\\
  \texttt{SPHERE$\_$SLOW}  & 10 & 19 & $\infty$ & 1000\\
  \texttt{TORUS$\_$FAST}  & 10 & 1.9 & 0.2 & 1000\\
  \texttt{COW$\_$SPHERE} & 10 & 0.8 & $\infty$ & 1300\\
  \texttt{COW$\_$TORUS} & 10 & 1.5 & 0.2 & 1300
 \enddata
\end{deluxetable}

The fiducial model \texttt{SPHERE$\_$FAST} assumes a spherical dust shell and a short rise-time of $t_{\rm rise} = 1.9$ d (e.g., similar to those which characterize FBOTs), thus placing it in the ``fast-rise'' regime $t_{\rm rise} \ll t_{\rm lc}$ discussed in Sec.~\ref{sec:fast_rise} and illustrated by the left panel of Fig.~\ref{fig:echo_qualitative}. The transient reaches its peak intrinsic luminosity of $L_{\rm pk} = 4\times 10^{42}$ erg s$^{-1}$ on a timescale $t_{\rm pk} \approx 7$ d. Model \texttt{SPHERE$\_$SLOW} is otherwise similar to the fiducial model, but considers a much slower-rising transient $t_{\rm rise} =19$ d, closer to those which characterize TDEs, and corresponding to the ``slow-rise'' regime $t_{\rm rise} \gg t_{\rm lc}$ discussed in Sec.~\ref{sec:slow_rise} (right panel of Fig.~\ref{fig:echo_qualitative}). The total dust shell mass $M = 0.15\,M_\odot$ results in an initial photosphere radius $R_{\rm ph,0} \approx 3\times 10^{16}$ cm and corresponding $L_{\rm thin} \approx 10^{42}$ erg s$^{-1}$ (Eq.~\eqref{eq:L_thin}).

The model \texttt{TORUS$\_$FAST} adopts the same fast-rising transient as the fiducial model, but assumes a torus-shaped dust envelope rather than a sphere, allowing for variations in the light-curve as a function of the observer's viewing angle relative to the torus axis. The torus density distribution follows Eq.~\eqref{eq:rhoangle} with $\Delta = 0.2$, normalized such that the total mass on the simulation domain is same as that of the spherical models.

\subsection{\texttt{SPHERE$\_$FAST}}

\begin{figure*}
    \centering
    \includegraphics[width=1.0\textwidth]{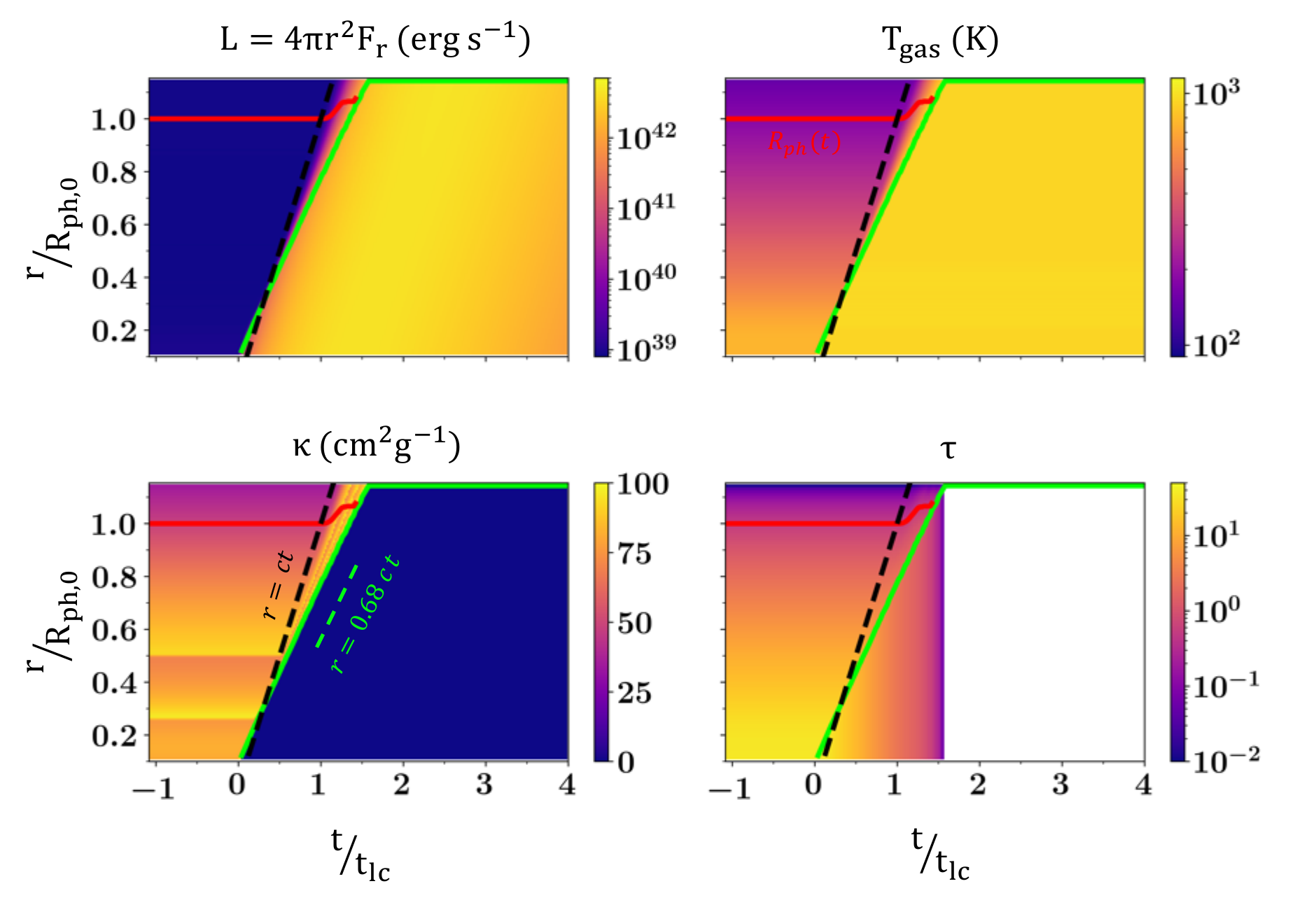}
    \caption{Radius-time diagrams showing (clockwise, starting from the top left) the evolution of the local radial luminosity $L = 4\pi r^{2}F_{r}$, gas temperature, $T_{\rm gas}$, optical depth to infinity $\tau \equiv \int_{r}^{\infty} \rho \kappa dr$, and Rosseland opacity $\kappa$. A red line depicts the photosphere ($\tau = 1$) while a green line depicts the sublimation front (minimum radius where $\kappa = 0$).}
    \label{fig:spacetime_fiducial}
\end{figure*}

Fig.~\ref{fig:spacetime_fiducial} shows how the radial profiles of key quantities evolve in time, including the radiation luminosity in the radial direction, $L = 4\pi r^{2}F_{\rm r}$, gas temperature, $T_{\rm gas}$, opacity, $\kappa$, and optical depth above a given radius, $\tau = \int_{r}^{\infty} \rho \kappa dr$ (Eq.~\eqref{eq:tau_r}). Asymmetries due to numerical errors were found to be insignificant, hence we report spherically-averaged values.

Prior to the start of the transient at $t = 0$, a radially-constant quiescent steady-state luminosity $L(r) = L_{\rm Q} = 10^{39}\,\text{erg}\,\text{s}^{-1} = const$ has been established through the dust/gas shell.  Consistent with that predicted by the steady-state diffusion equation for the assumed opacity law, the initial temperature decreases from $T_{\rm gas} \approx 800$ K near the inner boundary of the simulation to $\approx 200$ K at the photosphere. The radii at which individual species (olivine, troilite and refractary organics) sublimate are seen as discrete step-like jumps in the opacity.

The transient luminosity begins to rise at $t=0$, heating the gas to $T_{\rm gas} \gtrsim 10^{3}$ K and driving the sublimation front outwards in time, as depicted by a green solid line. The envelope is initially highly opaque to the reprocessed radiation $\tau(r = R_{\rm in}) \approx 35$, but the maximum $\tau (r_{\rm sub})$ to the center of the grid drops as the sublimation radius moves outwards. The latter moves outwards with a speed $v_{\rm sub} \approx 0.68 c$, close to light, as predicted in the fast-rise limit $t_{\rm rise} \ll t_{\rm lc}$ (though slightly lower than the analytic estimate 
$v_{\rm sub} \approx 0.72\,c$ one obtains from Eq.~\eqref{eq:vsub} for $t_{\rm rise} = 0.19t_{\rm lc}$). The sublimation front reaches the initial photosphere $r = R_{\rm ph,0}$, depicted by a solid red line, once the luminosity rises to the critical value $L \approx L_{\rm thin} \approx 10^{42}\,\text{erg}\,\text{s}^{-1}$ 
(Eq.~\eqref{eq:L_thin}) by the time $t \approx 1.35\,t_{\rm lc}$, or retarded time $t - t_{\rm lc} \approx 0.35t_{\rm lc}$ (close to the expected definition $t_{\rm rise} \approx t/2$; Eq.~\eqref{eq:t_rise}). Just prior to the arrival of the sublimation front at $r = R_{\rm ph,0}$, the instantaneous location of the photosphere moves outwards slightly to $r \approx 1.1\,R_{\rm ph,0}$ because of the modest rise in opacity with increasing gas temperature.

\begin{figure*}
    \centering
    \includegraphics[width=\textwidth]{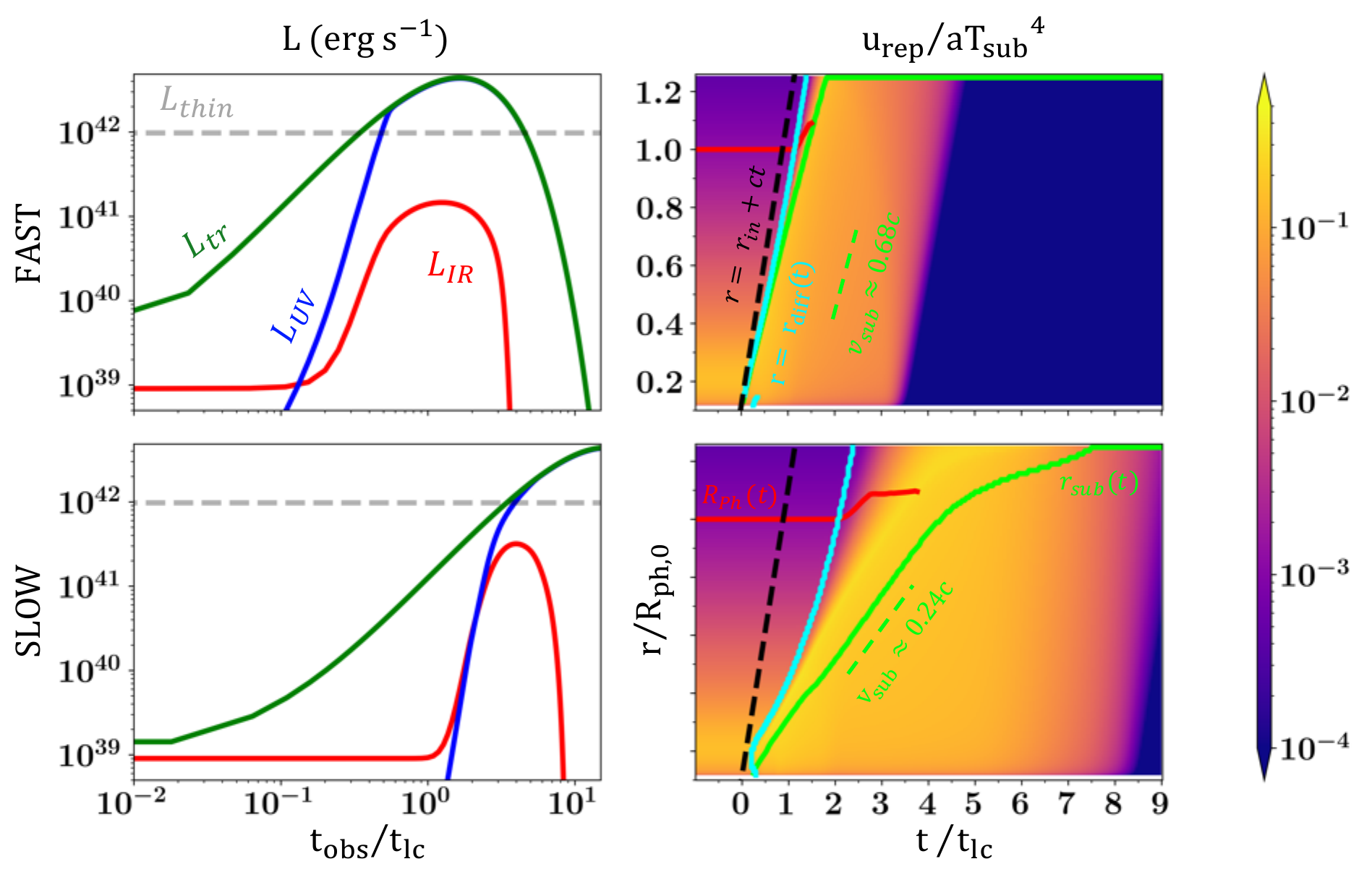}
    \caption{{\bf Left panels}: Light-curves from the fiducial fast-rise model \texttt{SPHERE$\_$FAST} (top)
    compared to the otherwise equivalent slow-rise model \texttt{SPHERE$\_$SLOW} (bottom).  We compare the intrinsic luminosity of the optical/UV transient $L_{\rm tr}$ (green lines; Eq.~\eqref{eq:lum_bdry_cond}) as injected at the center of our grid, to that which escapes through the external medium to a distant observer, $L_{\rm UV}$ (blue lines) as well as the reprocessed IR emission, $L_{\rm IR}$ (red lines). A horizontal dashed gray line shows the critical luminosity, $L_{\rm thin}$ (Eq.~\eqref{eq:L_thin}), at which dust is sublimated out to the initial photosphere radius. All times are measured with respect to the beginning of the transient (time of explosion), accounting for the light travel-time delay from the center of the grid to the measurement sphere at $1.2 R_{\rm ph,0}$ (see Eq.~\eqref{eq:t_obs}).
    {\bf Right panels}: Radius-time diagrams similar to Fig.~\ref{fig:spacetime_fiducial}, but showing the energy density of the non-radial reprocessed IR rays, $u_{\rm rep}$, normalized to the approximate critical value for dust sublimation, $a T_{\rm sub}^{4}$ (where $T_{\rm sub} = 1000\,K$ is the olivine
    sublimation temperature). The sublimation radius is shown in green, as in Fig.~\ref{fig:spacetime_fiducial}. A teal line shows the ``diffusion radius'' out to which reprocessed radiation from the transient has reached at a given time; the latter is defined as where $u_{\rm rep}$ first exceeds twice its initial steady-state value. Diffusion radius moves outwards faster than the sublimation front  
    for the slow-rising transient (bottom-right), hence $L_{\rm IR}$ rises above 
    the quiescent luminosity prior to the rise of $L_{\rm UV}$
    (bottom-left).}
    \label{fig:lightcurves_fast_slow}
\end{figure*}

Because the transient rises rapidly, the reprocessed radiation does not have time to diffuse far ahead of the sublimation front. This can be seen in the top right panel of Fig.~\ref{fig:lightcurves_fast_slow} showing the energy density of the reprocessed radiation; a teal line depicts the ``diffusion radius'', which we define as where the energy density locally first exceeds twice its initial value from the pre-transient state.  The diffusion radius closely follows the sublimation front, until both reach the photosphere where $\tau \sim 1$.

After the dust shell is completely sublimated, the transient luminosity continues to rise, peaking at $L_{\rm pk} \approx 4 \times  10^{42}\,\text{erg}\,\text{s}^{-1}$ (before decaying over a timescale $t_{\rm pk} \approx 0.7\, t_{\rm lc}$, see Table~\ref{tab:parameters_symbols}).
However, the gas does not continue to heat up further. After a given layer sublimates, its gas temperature freezes because the sudden drop in opacity for $T \gtrsim 10^{3}$ K (Fig.~\ref{fig:opacity_plot}) decouples the gas temperature from the rising radiation temperature.  

The top left panel of Fig.~\ref{fig:lightcurves_fast_slow} depicts the light-curves of the transient and its dust echo as seen by an observer at infinity, measured starting from the arrival time of the first photon released from the center of the grid at $t = 0$. The transient's luminosity, $L_{\rm tr}(t)$, as injected at the center of the grid (green line; Eq.~\eqref{eq:lum_bdry_cond}) begins to rise at $t = 0$. A blue line shows the escaping UV light of the transient, $L_{\rm UV}$, constructed from the (unabsorbed) radial rays. A red curve shows the reprocessed IR emission, $L_{\rm IR}$, calculated as described in Sec.~\ref{sec:light_curve_computation} from the non-radial rays.  A dashed gray line depicts an estimate of the critical luminosity $L_{\rm thin}$ (Eq.~\eqref{eq:L_thin}) at which the dust photosphere is sublimated.

For a brief interval after the explosion ($t - t_{\rm lc} \ll t_{\rm rise}$; $L \ll L_{\rm thin}$) the escaping luminosity is suppressed below the transient's intrinsic luminosity due to absorption by the dust shell. However, once the dust photosphere is destroyed by the expanding sublimation front at $t - t_{\rm lc} \sim t_{\rm rise}$ ($L \sim L_{\rm thin}$), both $L_{\rm UV}$ and $L_{\rm IR}$ begin to rise nearly simultaneously (consistent with expectations in the fast-rise scenario; Sec.~\ref{sec:fast_rise}). However, the bulk of the reprocessed IR radiation only reaches the observer over a few light-crossing times, as shown in the right top panel of Fig.~\ref{fig:lightcurves_fast_slow}. The reprocessed IR echo peaks at a luminosity $L_{\rm IR} \approx 0.1\,L_{\rm thin}$, consistent with analytic expectations (Eq.~\eqref{eq:LIR_fast}) given the fiducial model parameters $t_{\rm rise}/t_{\rm lc} \approx 0.2$ (Table.~\ref{tab:sim_param}). The shape of the light-curve near peak is approximately flat, albeit somewhat skewed to early times for reasons described below.

\subsection{\texttt{SPHERE$\_$SLOW}}

Model \texttt{SPHERE$\_$SLOW} considers a slower transient with $t_{\rm rise} \approx\, 2t_{\rm lc}$ encased in the same spherical dust shell as \texttt{SPHERE$\_$FAST} (Table~\ref{tab:sim_param}). We focus on describing the main differences between the two simulations, using Fig.~\ref{fig:lightcurves_fast_slow} as a guide.

The bottom-right panel of Fig.~\ref{fig:lightcurves_fast_slow} depicts the time-evolution of the radial profiles of reprocessed radiation energy density. Initially while still inside the photosphere, the sublimation front propagates outwards at $v_{\rm sub} \approx 0.24\,c$, close to the analytic estimate $\approx 0.21\,c$ based on Eq.~\eqref{eq:vsub}. The front reaches the initial photosphere radius at $t \approx 3.8\,t_{\rm lc}$, but by this time the photosphere has expanded slightly to $r \approx 1.1\,R_{\rm ph,0}$, similar to the fiducial model. The region ahead of the sublimation front becomes optically-thin to the reprocessed radiation, once $r_{\rm sub} \approx 0.9\,R_{\rm ph,0}$.
By contrast, in \texttt{SPHERE$\_$FAST}, this occurs only once $r_{\rm sub} \approx 1.07\,R_{\rm ph,0}$, i.e. the transition to optically-thin reprocessing 
coincided with the arrival of the sublimation front at the photosphere. The net energy of the radial-rays absorbed during this optically-thin portion of the evolution is $\approx 46\,\%$ of the total reprocessed energy throughout the simulation in \texttt{SPHERE$\_$SLOW}, but only $\approx 23\,\%$ in \texttt{SPHERE$\_$FAST}.

The sublimation front stalls upon reaching the photosphere, because a greater flux of radial-ray energy is required to sublimate the dust once the transient's radiation begins to leak out. The duration over which reprocessing occurs on a given radial scale can be quantified by the distance separating the diffusion radius (teal curve) and the sublimation radius (lime curve). In particular, we see that reprocessing near the photosphere ($r \approx R_{\rm ph,0}$) lasts for $\approx 2\,t_{\rm lc} \approx t_{\rm rise}$, as expected. The peak reprocessed energy density achieved throughout the grid across all times (the brightest regions in Fig.~\ref{fig:lightcurves_fast_slow}) correspond to $u_{\rm rep, pk} \approx 0.4\, aT_{\rm sub}^4$ for \texttt{SPHERE$\_$SLOW}
 compared to $u_{\rm{rep,pk}} \approx 0.26\,aT_{\rm sub}^4$ for \texttt{SPHERE$\_$FAST}. 
  
 The bottom-left panel of Fig.~\ref{fig:lightcurves_fast_slow} shows that both the escaping UV and reprocessed IR light-curves begin to rise above the quiescent luminosity after a delay $t_{\rm on} \approx 2\,t_{\rm lc}$ following the onset of the transient (i.e., the initial rise of $L_{\rm tr}$), in rough agreement with the estimate $t_{\rm on} \approx 2.2\,t_{\rm lc}$ based on Eq.~\eqref{eq:t_esc}. The IR light-curve also peaks on the analytically estimated timescale $t_{\rm IR} \approx 2\,t_{\rm lc} \approx t_{\rm rise}$ (Table~\ref{tab:timescales_luminosities}) albeit at a luminosity $L_{\rm IR} \approx 0.3\,L_{\rm thin}$, somewhat lower than the analytic estimate. As the IR light-curve rises to its peak, the UV and IR light-curves closely track each other, since the relative delays in the arrival time of reprocessed photons are insignificant in the slow-rise limit, i.e. the dust does not ``echo".

\begin{figure}
    \centering   \includegraphics[width=0.48\textwidth]{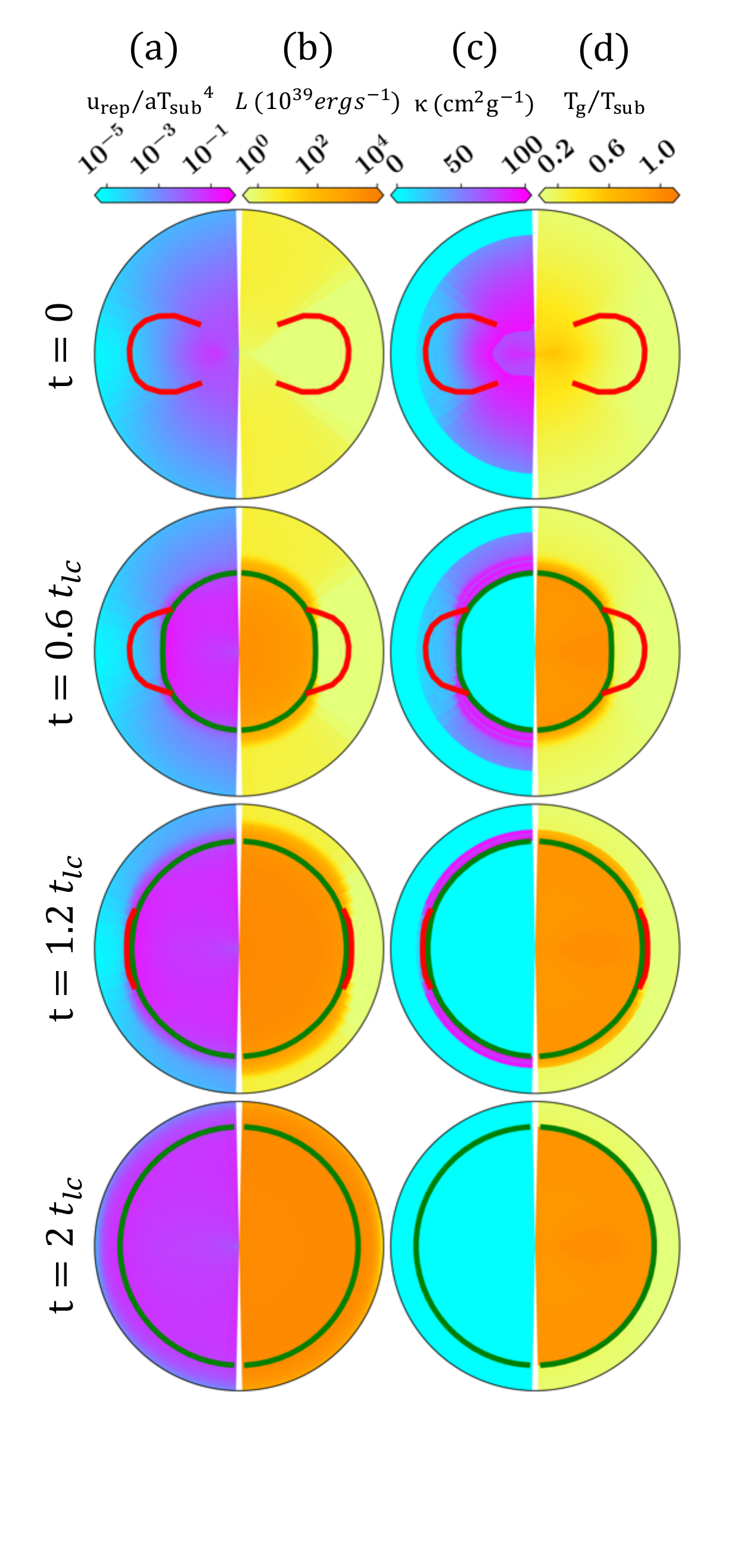}
    \caption{Two-dimensional ($r,\theta$) snapshots of the model \texttt{TORUS$\_$FAST}, taken at four times after the onset of the transient as marked. We show the reprocessed IR radiation energy density $u_{\rm rep}$
    (panel (a)), luminosity $L = 4\pi r^2 F_r$ (panel (b)), opacity (panel (c)), and gas temperature normalized to $T_{\rm sub} = 1000$ K (panel (d)). We calculate $u_{\rm rep}$ from the non-radial rays (see Sec.~\ref{sec:light_curve_computation}) and normalize its value to $aT_{\rm sub}^4$. A green contour depicts the sublimation surface $r_{\rm sub}(t,\theta)$ inside which $\kappa = 0$, while a red contour marks the photosphere of the radial rays $\tau(t,\theta) = 1$. We show radii $0.1\,R_{\rm ph,0} < r < 2 R_{\rm ph,0}$ in logarithmic scale.}
    \label{fig:torus_snapshots}
\end{figure}

\begin{figure*}
    \centering
    \includegraphics[width=\textwidth]{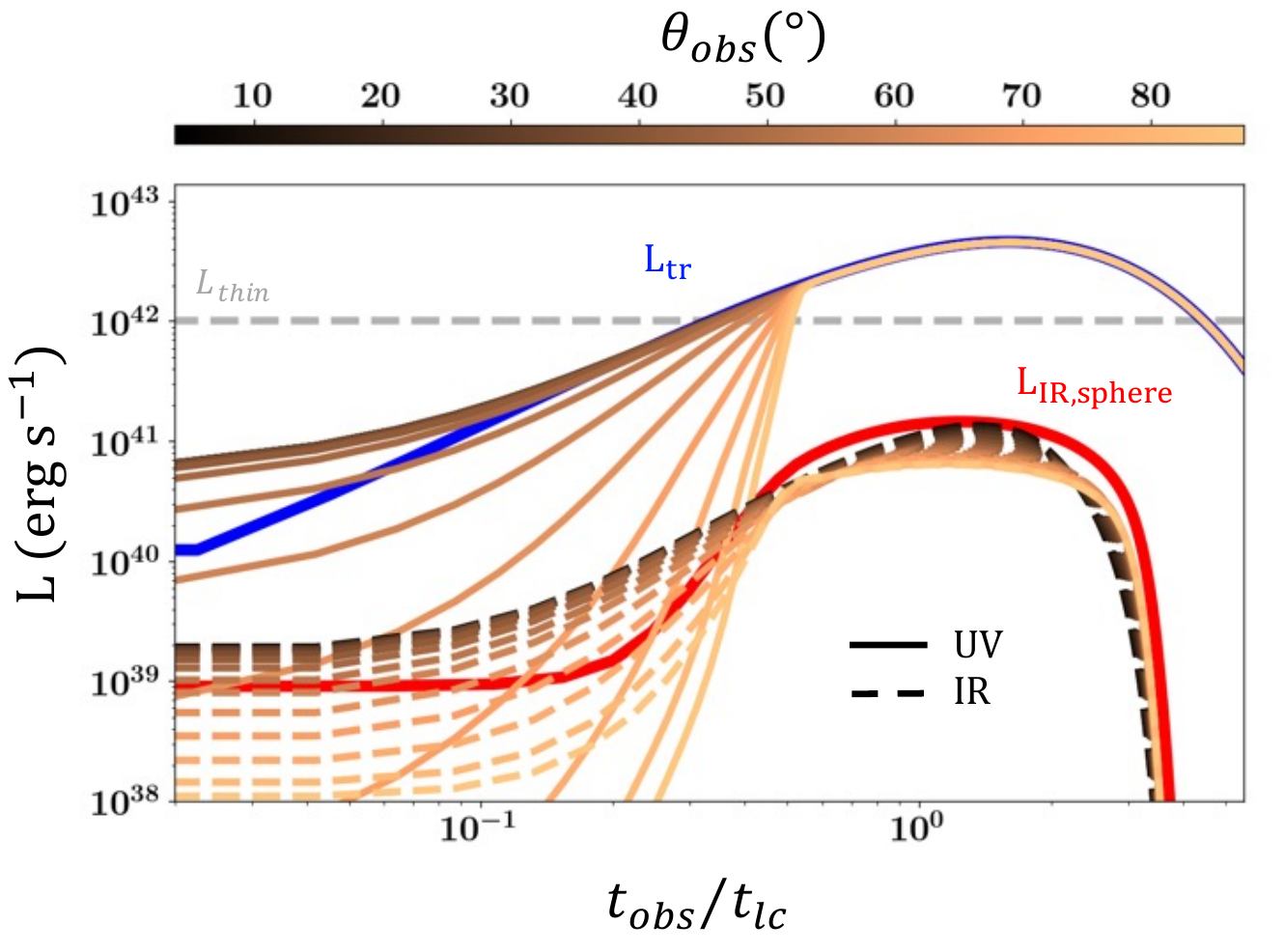}
    \caption{Isotropic luminosities 
    $L = 4\pi r^2 F$ for \texttt{TORUS$\_$FAST} (covering angle
    $\Delta = 0.2$), color
    indexed according to observer angle $\theta_{\rm obs}$,
    as measured from the symmetry axis, i.e. $\theta_{\rm obs} = 0$ corresponds to an observer
    viewing the system from the pole. (Escaping) UV and 
    IR light-curves are plotted in solid and dashed lines,
    respectively. Blue solid line shows the transient 
    luminosity injected at the center of the grid,
     as in Fig.~\ref{fig:lightcurves_fast_slow}. 
     In comparison,
    the IR light-curve of the ``equivalent" spherical model
    \texttt{SPHERE$\_$FAST} is shown in red.}
    \label{fig:torus_fast_lc}
\end{figure*}

\subsection{\texttt{TORUS$\_$FAST}}
\label{sec:torus_fast}

\begin{figure*}
    \centering
    \includegraphics[width= \textwidth]{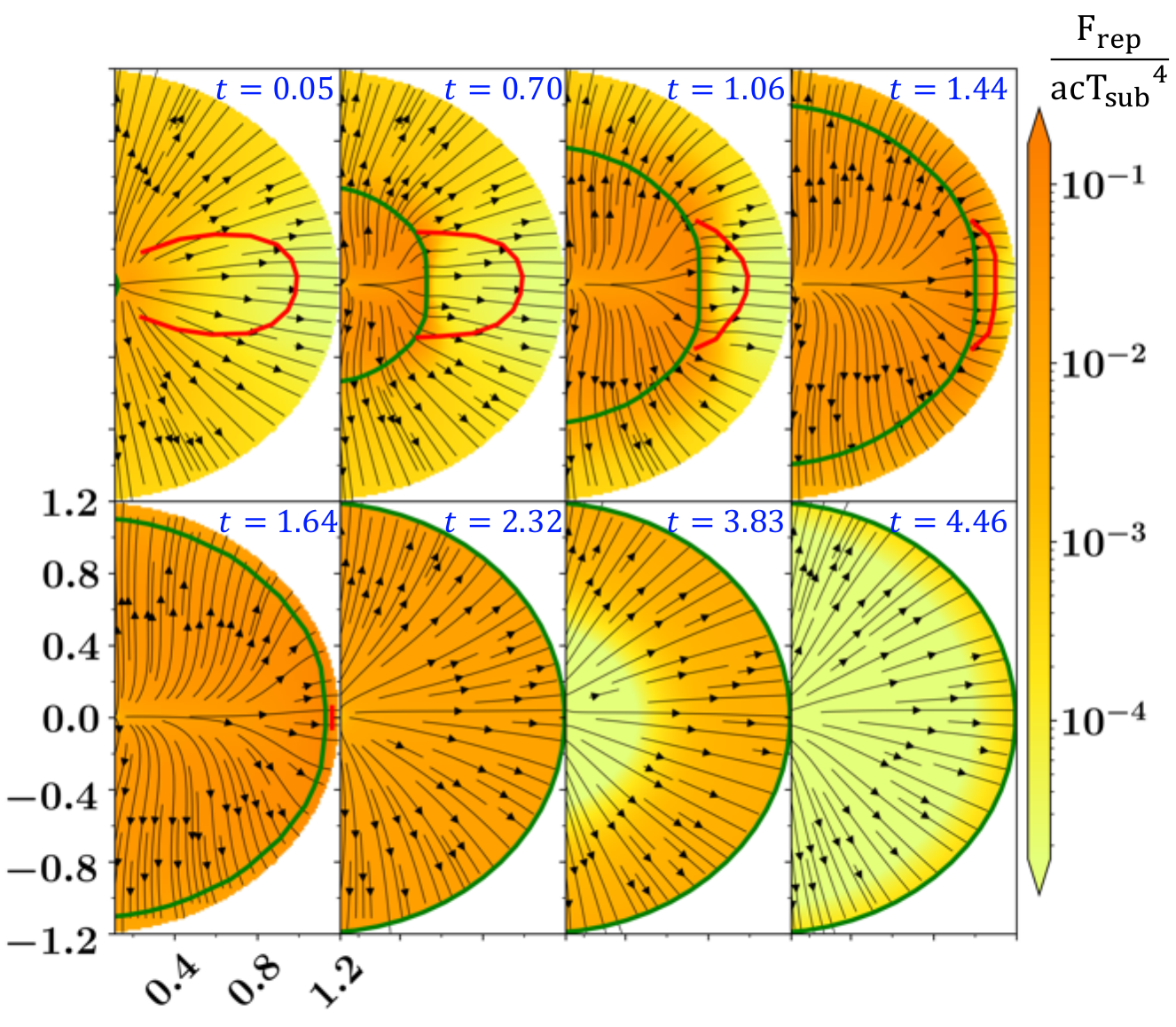}
    \caption{Snapshots of \texttt{TORUS$\_$FAST} showing reprocessed energy flux
    (calculated using non-radial rays). Blue labels
    mark the time after explosion 
    in units of $t_{\rm lc}$.
    The color plot shows the magnitude of the energy flux scaled to $acT_{\rm sub}^4$, with $T_{\rm sub} = 10^3\,K$, while streamlines show the direction of the flux.  Green curve shows the sublimation surface, and red curve shows $\tau = 1$ surface along $\theta = \text{const}$, if it exists. Distances are in units of $R_{\rm{ph, 0}}$. Reprocessed 
    energy flux is directed towards the poles until the complete destruction of the dust sphere at $t = 1.64\,t_{\rm lc}$.}
    \label{fig:torus_rep_flux}
\end{figure*}

The \texttt{TORUS$\_$FAST} model takes the dust shell to be torus-shaped rather than spherical, otherwise fixing the transient properties the same as the spherical model \texttt{SPHERE$\_$FAST}. Figure \ref{fig:torus_snapshots} shows temporal snapshots of two-dimensional $(r,\theta)$ maps of several quantities: the energy density of the
non-radial (reprocessed) rays (panel a), radial luminosity $L = 4\pi r^2 F_r$ (panel b), opacity (panel c), and gas temperature (panel d). 

As in the \texttt{SPHERE} cases, the constant luminosity $L_{\rm Q} \approx 10^{39}\,\text{erg}\,\text{s}^{-1}$ injected isotropically from the center of the grid prior to the rise of the transient establishes a roughly radially-constant luminosity
$L = 4\pi r^2 F_r$ across the simulation domain (panel (b), $t = 0$). The radiation flux along the polar directions slightly exceeds that in the optically-thick equatorial region because the initially equatorially-directed radiation diffuses towards the polar directions due to their lower optical depth. Near the equator, the gas temperature reaches $\approx 800\,$K at small radii, dropping to $\approx 200 K$ outside the  
photosphere (the photosphere surface is denoted by a red contour). The mean opacity (panel c) peaks at $\kappa \approx 100\,\text{cm}^2\,\text{g}^{-1}$ in the hottest regions where $T_g \approx 800 $K. 

By the second snapshot at $t = 0.6\,t_{\rm lc}$ the transient begins to rise, and the sublimation surface $r_{\rm sub}(t, \theta)$ (denoted by a green contour in Fig.~\ref{fig:torus_snapshots}) begins moving outwards. The sublimation surface maintains a roughly spherical shape during its expansion because in the fast-rise limit its location is mainly determined by
the luminosity of the transient at the retarded time (see Sec.~\ref{sec:analytical_estimates}). In the equatorial regions just upstream
of the sublimation front, sharp
radial gradients are seen in the gas temperature, luminosity and opacity.

By the third snapshot ($t \approx t_{\rm lc}$),
the transient luminosity reaches $L \approx L_{\rm thin} \approx 10^{42}\,\text{erg}\,\text{s}^{-1}$. The now-thin dust shell near the photosphere
continues to reprocess radiation until being 
completely destroyed by the rising temperature. The region inside the sublimation front ($\kappa = 0$) is full of reprocessed rays propagating around freely, reaching their peak energy density $u_{\rm rep} \approx 0.3\,aT_{\rm sub}^4$. These rays are also free to escape to an external observer without encountering significant optical depth after this stage in the transient rise.

The final snapshot in Fig.~\ref{fig:torus_snapshots} depicts a moment $t = 2\,t_{\rm lc}$, after complete dust destruction. The gas temperature within the sublimation front has frozen at $T_g \approx 10^3\,$K. The reprocessed radiation gradually leaves the domain, resulting in a drop in the reprocessed energy density (darker purple regions in panel (a)).

Fig.~\ref{fig:torus_fast_lc} shows the UV (solid) and IR (dashed) light-curves for \texttt{TORUS$\_$FAST}
for various viewing angles angles (colored according
to $\theta_{\rm obs}$), in comparison to the intrinsic transient luminosity $L_{\rm tr}$ (blue line). Also for comparison with a red line is the IR light-curve of the (otherwise equivalent) spherical-dust shell model \texttt{SPHERE$\_$FAST} (Fig.~\ref{fig:lightcurves_fast_slow}). 
The qualitative features of the IR light-curve around
the peak luminosity are similar for all observing angles.  However, we observe an overall trend towards an earlier rise,
higher peak luminosity,
and shorter duration spent near the peak, for viewers located closer to the poles. The IR light-curves seen by equatorial viewers $\theta_{\rm obs} \approx 90^\circ$ (brighter dashed lines) also more closely resemble those of the spherical model \texttt{SPHERE$\_$FAST}, albeit scaled down in magnitude $L_{\rm IR, torus} \approx 0.4\,L_{\rm IR, sphere}$.

\begin{figure*}
    \centering
    \includegraphics[width= \textwidth]{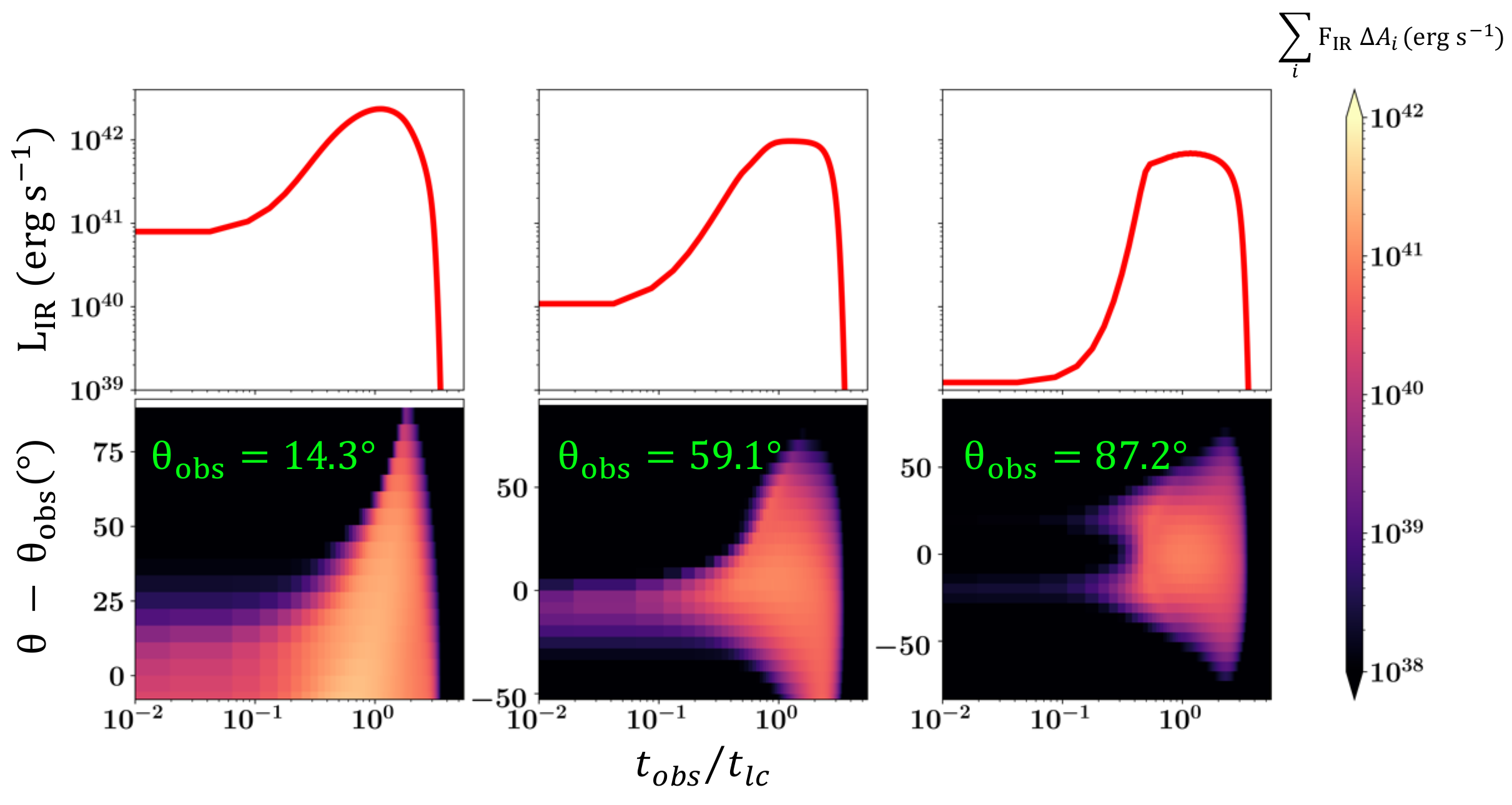}
    \caption{\textbf{Top panel:} IR light-curves for observers with viewing angles $\theta_{\rm obs} = 14.3$, $\theta_{\rm obs} = 59.1$ an $\theta_{\rm obs} = 87.2$ degrees
    measured from the torus polar axis. \textbf{Bottom panel:} The
    sum of reprocessed energy flux $F_{\rm IR}$ over area elements $A_i$ on the sphere, 
    that contribute to the light-curve at a given observation time and emission angle $\theta$. Here $\theta$ denotes the latitude of the area element on the spherical surface used in the light-curve computation (Sec.~\ref{sec:light_curve_computation}).}
    \label{fig:torus_lc_contributions}
\end{figure*}

The earlier rise and larger peak luminosity seen by polar observers $\theta_{\rm obs} \approx 0$ can be understood as a consequence of
reprocessed radiation energy flux
being directed preferentially towards the poles where the optical depth is lower. Fig.~\ref{fig:torus_rep_flux} shows the
direction and magnitude of reprocessed energy flux (computed from non-radial rays)
at key moments during the simulation. By $t = 1.44\,t_{\rm lc}$ the energy flux in the polar regions has already risen by an order of magnitude, whereas there is still an optical depth $\tau \approx 1$ 
ahead of the reprocessed rays in the equatorial direction (red contour), delaying the rise at the equator (bright yellow region). 
After $t \approx 1.64\,t_{\rm lc}$ even the equator has become optically-thin to radial rays, and dust is completely destroyed. After this moment, the domain is gradually depleted from the reprocessed rays, and the flux becomes more isotropic.

Fig.~\ref{fig:torus_lc_contributions} further explores how variations in the IR light-curve with the observer angle $\theta_{\rm obs}$ arises due to the geometry of the reprocessing.  We present a ``latitude-time" diagram in which the intensity of the color indicates, at each point in time, the contribution to the total flux received from a given latitude $\theta$ of the spherical surface used in our light-curve calculation (Sec.~\ref{sec:light_curve_computation}). Every observer receives light from their respective hemispheres. At the initial onset of transient, the polar observer ($\theta_{\rm obs} = 14.3^\circ$) is already receiving an isotropic luminosity $L = 4\pi r^2 F \approx 10^{41}\,\text{erg}\,\text{s}^{-1}$ from latitudes $ 10^\circ \lesssim \theta \lesssim 40^\circ$,
which is mainly the escaping reprocessed energy flux depicted in 
Fig.~\ref{fig:torus_rep_flux}. By $t\approx 0.2\, t_{\rm lc} \approx t_{\rm rise}$, the light-curves seen by all observing angles begin to rise. Observers located close to the pole ($\theta_{\rm obs} = 14.3^\circ$, $59.1^\circ$) first receive direct light from $\theta \approx \theta_{\rm obs}$. By contrast, due to the higher optical depth through the equatorial regions of the torus, the first light seen by equatorial observers ($\theta_{\rm obs} = 87.2^\circ$) is indirect reprocessed radiation received from higher latitudes $\theta \leq 70^\circ$, where the dust is sublimated earlier. Reprocessed emission from the equator $\theta \approx 90^\circ$ only contributes to the light-curve after $t \approx 0.5\,t_{\rm lc}$ once the equator becomes optically-thin to reprocessed rays (see also Fig.~\ref{fig:torus_snapshots}). These rays are also
the ones that carry most of the energy, since they were created when $L \approx L_{\rm thin}$. Therefore, the light-curve of the equatorial observer 
$\theta_{\rm obs} \approx 87.2^\circ$ is flatter in its subsequent evolution than for polar observers. On the other hand, the polar observer only receives the rays reprocessed at $L \approx L_{\rm thin}$ (from the equatorial region) after a light crossing time $t \approx t_{\rm lc}$, thus causing the polar light-curve to be skewed to later times. 

\begin{figure*}
    \centering
    \includegraphics[width=\textwidth]{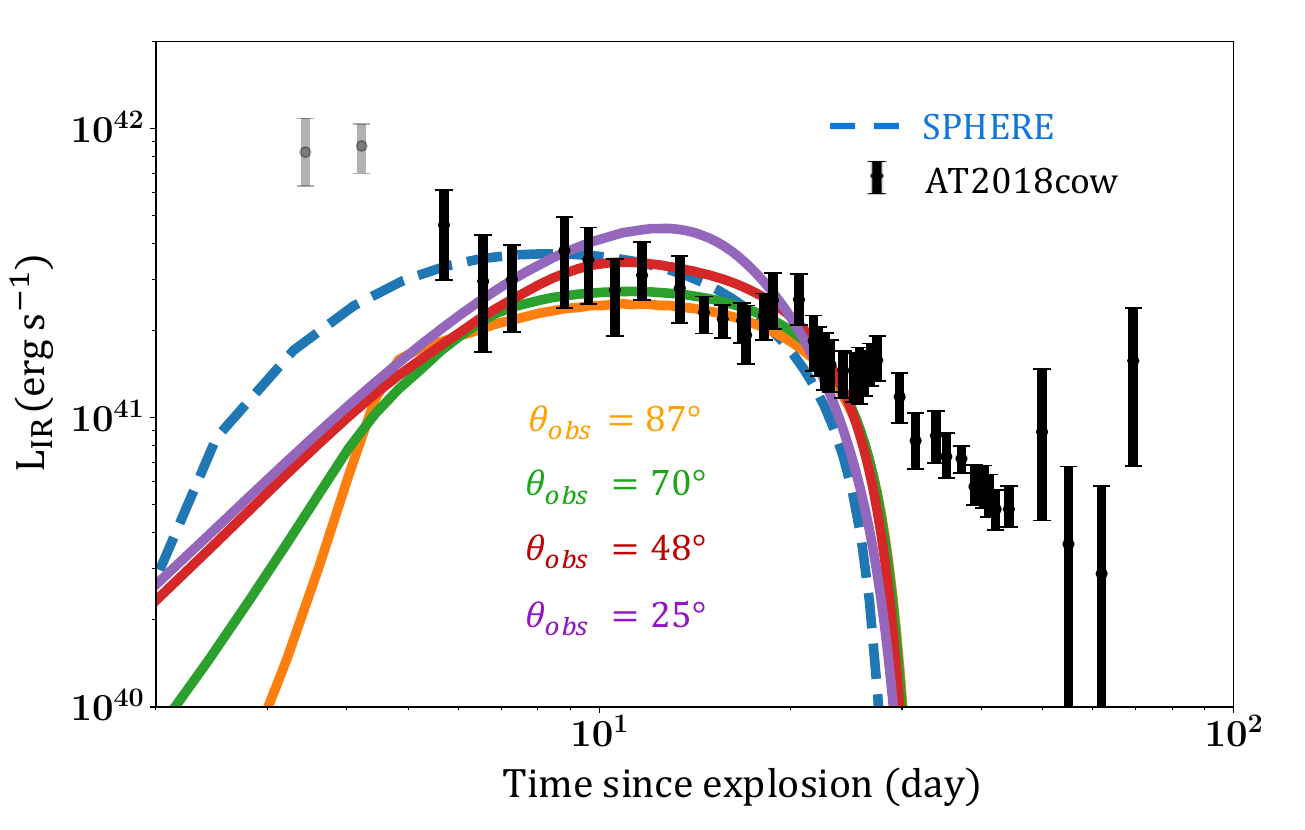}
    \caption{IR light curves of models $\texttt{COW$\_$SPHERE}$ 
    (blue dashed line) and $\texttt{COW$\_$TORUS}$ (solid lines),
    compared to the observed IR excess of AT2018cow \citep{Perley+19} for different observer viewing angles. We do not attempt to fit the first two data points (shown in gray) since the IR excess does not yet clearly stick out from the optical transient spectrum at these early epochs. For the torus-shaped CSM model $\texttt{COW$\_$TORUS}$, equatorial observers 
    $\theta_{\rm obs} = 87^\circ$ (orange line), $\theta_{\rm obs} = 70^\circ$  (green line) provide a better fit to the data than for pole-on observers. The spherical model fits the data point at $t \approx 5\,d$ better than the torus model, since the spherical model rises to its peak faster (see Table ~\ref{tab:sim_param}). The \texttt{COW$\_$TORUS} fit would improve if we had adopted a higher sublimation temperatures $T_{\rm sub} \geq 1700\,K$ (as necessary to explain the IR spectra) and correspondingly shorter $t_{\rm rise}$, which however we are not able to probe due to numerical limitations (see discussion in Sec.~\ref{sec:AT2018cow}).}
    \label{fig:cow_fit}
\end{figure*}

\section{Application to AT2018cow}
\label{sec:AT2018cow}

As an application of our results, we present two models which attempt to reproduce the excess IR emission seen in AT2018cow and interpreted as a dust echo by \citet{Metzger&Perley23}. The first model, \texttt{COW$\_$SPHERE}, assumes a spherical dust shell and other parameters (Table~\ref{tab:sim_param}) similar to \texttt{SPHERE$\_$FAST} (Table~\ref{tab:parameters_symbols}), except for a slightly shorter transient rise-time $t_{\rm rise} \approx\, 0.8\,d$. The second model, \texttt{COW$\_$TORUS}, assumes a torus-shaped dust shell (equatorial opening-angle $\Delta = 0.2$), and a longer $t_{\rm rise} = 1.5\,$d than \texttt{SPHERE$\_$FAST}.

A more significant change for these simulations is that we artificially increase the sublimation temperature of the limiting species (olivine) from $T_{\rm sub} \approx 1000\,K$ to $1300\,$K. The silicate sublimation temperature $T_{\rm sub} \approx 1000\,K$ assumed in our opacity law \citep{Pollack+94} are substantially lower than those found by \citet{Draine&Lee84}, in part as a result of the optical constants assumed. Empirically, a high dust temperature $T_{\rm d} \sim 2000\,$K is required to explain the spectral peak of the IR excess in AT2018cow \citep{Perley+19,Metzger&Perley23}. More generally, dust echo models for AT2018cow are inconsistent for $T_{\rm sub} \lesssim 1300\,K$, because while the IR emission requires a ``fast-rising" model, the observed luminosity requires $t_{\rm rise} \geq t_{\rm lc}$ for $T_{\rm sub} \lesssim 1300\, K$ according to Eq.~\eqref{eq:LIR_fast} (see the discussion surrounding Eq.~\ref{eq:t_lc_lower_limit} below). An increase in the sublimation temperature can also be justified if carbonaceous grains (with higher sublimation temperatures than silicates) were present in the environment of AT2018cow \citep{Metzger&Perley23}. The reason we adopt $T_{\rm sub} = 1300\,K$ instead of an even higher value $T_{\rm sub} \geq 1700\,K$ the result of numerical limitations: we found that simulating fast-rising transients $t_{\rm rise} \ll t_{\rm lc}$ in our setup requires a prohibitively high spatial resolution. Resolving the transient rise at the domain inner boundary (Eq.~\ref{eq:lum_bdry_cond}) requires a spatial resolution $\sim c\,t_{\rm rise}$ on radial scales $\sim R_{\rm ph,0} = c\,t_{\rm lc}$.

Fig.~\ref{fig:cow_fit} shows the reprocessed light-curves obtained
for the two models in comparison to light-curve data on AT2018cow from \citet{Perley+19}, \citet{Metzger&Perley23}. The IR excess from AT2018COW lasted for over a month, a period that would likely need to include an extended phase of emission from optically-thin dust well above the original dust photosphere (see Appendix \ref{sec:opticallythin}). Since our light-curve calculations only address the reprocessed emission from the optically-thick dust on small radial scales around the transient (Sec.~\ref{sec:opacity_cutoff}), we compare our models only to the early emission phase of duration
$t_{\rm IR} \approx 2\,t_{\rm lc} \approx 20\,\text{days}$.  The spherically symmetric model 
\texttt{COW$\_$SPHERE} provides a reasonably good fit to data, except for the earliest two data points (however, at these first two epochs the IR excess has not yet clearly emerged from the hotter transient spectrum, rendering the start-time of the putative echo uncertain; see \citealt{Metzger&Perley23}).

Models for the multi-wavelength emission from AT2018cow point to a highly aspherical environment surrounding the explosion, in which most of the mass is concentrated in the equatorial regions away from the polar axis of the fastest outflowing gas (e.g., \citealt{Margutti+19,Metzger2022}). In a scenario where the IR radiation arises due to reprocessing by an optically-thick torus, the light-curve shape demands particular combinations of the rise time, observer angle and degree of asymmetry ($\Delta$) for aspherical models. In particular, highly aspherical distributions with $\Delta \ll 1$ seem to provide poorer fit to data for the following considerations. Informed by the light-curves of \texttt{TORUS$\_$FAST} in Fig.~\ref{fig:torus_fast_lc} and the surrounding discussion (Sec.~\ref{sec:torus_fast}), we note that equatorial viewing angles with 
$\theta_{\rm obs} \approx 90^\circ$ light-curves are flatter and last longer near the
peak, thus providing a better fit to the light-curve shape of AT2018cow.  Models with $\theta_{\rm obs} \approx 90^\circ$ and $\Delta \ll 1$
achieve lower peak IR luminosities compared to otherwise equivalent spherical models, hence requiring longer rise times $t_{\rm rise}$ near $L \approx L_{\rm thin}$ compared to spherical models to match AT2018Cow. However, the peak timescale of the transient $t_{\rm pk} \approx 3\,\text{d}$ together with the peak UV and luminosities $L_{\rm pk} \approx 10^{44}\,\text{erg}\,\text{s}^{-1}$, $L_{\rm{IR, pk}} \approx 10^{42}\,\text{erg}\,\text{s}^{-1}$ are in tension with ``slow-rise limit'' (Sec.~\ref{sec:analytical_estimates}) 
$t_{\rm rise} \gtrsim t_{\rm lc}$, since 
$L_{\rm{IR, pk}} \leq L_{\rm thin}$ requires
\begin{align}
     t_{\rm lc} \gtrsim 3\,\text{day}\,
     \left(\frac{L_{\rm IR}}{3\,\times\,10^{41}\,\text{erg}\,\text{s}^{-1}}\right)^{1/2}\left(\frac{T_{\rm sub}}{1300\,K}\right)^{-2} 
     \label{eq:t_lc_lower_limit}
\end{align}
For $\Delta = 0.2$, the peak luminosities for an equatorial viewer $\theta_{\rm obs} = 90^\circ$ are a factor $\approx 2$ (see also Sec.~\ref{sec:torus_fast}) below the equivalent spherical model, thus enabling us to fit the torus model 
 $\texttt{COW$\_$TORUS}$, provided we take $t_{\rm rise} = 1.5\,d$ and hence stay in the fast-rise regime. We expect this fit to become poorer for more aspherical dust distributions.
 
\section{Conclusions}
\label{sec:conclusions}

We have presented 2D axisymmetric, time-dependent radiation transport simulations of transients enshrouded by dust.  We focus on situations in which the enshrouding dust is initially optically-thick to the transient's light, but which become optically-thin later as the dust shell is progressively heated and sublimated to larger radii by the rising luminosity. Using the intensity field self-consistently computed
by the radiation transport module of \texttt{Athena++}
\citep{Jiang21}, we constructed bolometric light-curves of radiation reprocessed by dust as it appears to an external observer at infinity. Our setup is relatively general, and can be applied to a wide class of transients ranging from TDEs to FBOTs. 

Transients enshrouded by dust can be classified into two broad categories as ``fast-rising'' vs. ``slow-rising'', depending on how quickly the light-curve rises compared to the light crossing time of the dust photosphere (Fig.~\ref{fig:transients_parameterspace}).  The IR light-curves exhibit distinct characteristic time-and luminosity scales in these two regimes (Fig.~\ref{fig:echo_qualitative}, Table~\ref{tab:timescales_luminosities}). For example, for slow-rising transients the IR light-curve begins to rise on an intermediate timescale between $t_{\rm rise}$, $t_{\rm lc}$, once the diffusion speed of the reprocessed radiation through the dust envelope exceeds the sublimation front velocity, depending on the density profile of the medium (Fig.~\ref{fig:echo_qualitative}, Eq.~\eqref{eq:t_esc}). 

Motivated by FBOTs and TDEs, we explored transients surrounded by aspherical dust distributions (Sec.~\ref{sec:torus_fast}, Sec.~\ref{sec:AT2018cow}). The light-curves in this case depend on the observer's viewing angle relative to the dust symmetry axis. For dust concentrated within a torus of characteristic equatorial solid angle $\approx 0.4\pi$ ($\Delta = 0.2$, Eq.~\eqref{eq:rhoangle}) the light curve's qualitative behavior is insensitive to viewing angle, with differences in peak luminosity remaining within a factor of $\lesssim 2$. However, in detail the shape and symmetry of the IR echo, and its arrival relative to the direct UV light of the transient, differs for polar and equatorial viewers.  As a proof of concept, we presented two models$-$with and without spherical symmetry$-$in reasonable agreement with the early IR excess emission observed from AT2018cow, thus supporting the general model of \citet{Metzger&Perley23}. Fitting AT2018cow required adopting a higher sublimation temperature  $T_{\rm sub} \approx 1300\,K$ then our opacity table assumes \cite{Pollack+94}, similar results would be obtained if, e.g. different grain species with higher sublimation temperatures were considered.  Indeed, dust grain temperatures $T \approx 1700-2000$ K were found necessary to explain the spectrum of the infrared excess \citep{Metzger&Perley23}, though we were only able to explore models with $T_{\rm sub} = 1300\,K$ due to numerical limitations.

Perhaps the biggest drawback of our treatment is the use of a grey approximation regarding the dust opacities (Sec.~\ref{sec:opacity}).  As a consequence, our calculations underestimate the amount of absorption (which mainly occurs in UV wavelengths) since the mean opacities we use at the relevant gas temperatures $T \lesssim 2000\,K$ are representative of the
dust opacity in IR wavelengths, which can be a factor $\sim 3-10$ smaller than UV opacities, depending on the dust grain sizes \citep{Birnstiel+18}. 
However, from the point of view of bolometric light-curves, 
the overall magnitude of opacity is largely degenerate with other parameters of the environment, such as the density normalization, metallicity and size of the dust grains. Interpreted within this framework, our method should be sufficient to capture the bulk light-curve behavior such as total amount of absorbed energy and the timescales over which the reprocessing occurs. On the other hand, there are several limitations to our method which would require a multi-frequency group calculation to overcome. Most notably, frequency-averaged dust opacities underestimate the true opacity to UV light in optically-thin regions even more severely at large radii where gas is even colder; this motivated our decision to restrict our calculation to reprocessing from optically-thick regions (Sec.~\ref{sec:opacity_cutoff}). Scattering effects are also not captured by our method, since we are required to distinguish direct UV from reprocessed IR rays by their direction (Sec.~\ref{sec:resolution}).

When dust is present, it dominates over other sources of opacity, justifying our neglect thereof (Sec.~\ref{sec:opacity}). Our results would change qualitatively, if gas at small scales cleared from dust were to become opaque to the transient due to, e.g., electron scattering, prior to complete destruction of the dust at larger scales.  In this case, the central transient light would need to propagate through this newly formed inner photosphere, potentially slowing down the propagation of the sublimation front at larger scales, and pushing, e.g. a fast-rising transient to the slow-rising regime.

\acknowledgements

We thank Kishalay De for helpful discussions and Zhaohuan Zhu for providing tabulated dust opacities.  S.~T.~and B.~D.~M.~are supported in part by NASA (grant numbers 80NSSC22K0807 and 80NSSC24K0408), the Simons Foundation (grant number 727700), and the National Science Foundation (grant number AST-2009255).  The Flatiron Institute is supported by the Simons Foundation.

\appendix

\section{Optically-thin reprocessing}
\label{sec:opticallythin}

Even once dust has been sublimated out to the photosphere, $r_{\rm sub} > R_{\rm ph,0},$ optically-thin dust from larger radii $r > R_{\rm ph,0}$ can continue to reprocess the transient's growing luminosity.  There are in principle two phases of this optically-thin dust emission. 

The first phase occurs as the transient luminosity continues to rise from $L_{\rm thin}$ to its peak luminosity $L_{\rm pk}$, and the optical depth of the dust continues to drop as $\tau \propto r_{\rm sub}^{1-p} \propto L_{\rm tr}^{(1-p)/2}$ due to sublimation.  Hence the energy absorbed by the dust over the timescale $\sim t$ scales as: 
\be E_{\rm abs} \propto (L_{\rm tr}\cdot t)\cdot \tau \propto t \cdot L_{\rm tr}^{(3-p)/2}. \ee
In the fast-rise limit ($t_{\rm lc} \gg t_{\rm pk}$, though defined this time relative to the peak-time of the transient) and assuming a generic transient power-law rise-law $L_{\rm tr} \propto t^{\beta}$ (e.g, Eq.~\eqref{eq:LUVrise}), this re-radiated energy will reach an observer on a timescale $t_{\rm lc} \propto 2r_{\rm sub}/c \propto L_{\rm tr}^{1/2} \propto t^{2/\beta}$. The NIR echo light-curve from these optically-thin layer thus evolves in time as:
\be
L_{\rm IR, thin,1} \propto \frac{E_{\rm abs}}{t_{\rm lc}} \propto t^{\frac{(2-p)\beta + 2}{2\beta}} \underset{\beta = 2}\propto  t^{\frac{3-p}{2}},\,\,\, \tau < 1.
\ee
This emission will dominate the emission of the earlier optically-thick echo for sufficiently shallow density profiles:
\be
p \le \frac{2}{\beta}+2 \underset{\beta = 2}=3.
\ee
For such steeply decaying density profiles satisfying this expression, $L_{\rm IR,thin,1}(t)$ decreases in time, and it will only come to dominate over the (approximately flat) echo emitted by the optically-thick dust emission after the latter has finished.

A second phase of the optically-thin dust echo occurs after the transient has reached its peak luminosity and the sublimation radius has saturated at its maximum value $r_{\rm sub,max} \equiv r_{\rm sub}(L_{\rm pk}$) such that $\tau$ stops decreasing. An IR echo continues to later times, as a fraction $\propto \tau \propto r^{1-p}$ of the total energy radiated by the transient ($E_{\rm rep} \sim L_{\rm pk}t_{\rm pk}$), is reprocessed by dust at yet larger radii $r \gg r_{\rm sub,max}$, at correspondingly later times, $t \approx t_{\rm lc} \sim 2r/c$. The luminosity emitted by these never-sublimated layers decays as: 
\be
L_{\rm IR, thin,2} \propto \frac{E_{\rm tot}\tau}{t_{\rm lc}} \propto r^{-p} \propto t^{-p}
\ee
Unlike the earlier NIR optically-thin echo produced mainly by gas near the sublimation temperature, the temperature of this cooler-emitting dust will decay, roughly as
\be
T \propto \left(\frac{L_{\rm IR,thin,2}}{r^{2}}\right)^{1/4} \propto t^{-\frac{2+p}{4}},
\ee
i.e. the characteristic wavelength of the emission will move further into the IR with time.  

\bibliographystyle{aasjournal}
\bibliography{refs,refs2}

\end{CJK*}

\end{document}